\pdfoutput=1
\documentclass[preprint,aps,prstab]{revtex4}

\usepackage{graphicx}
\usepackage{dcolumn}
\usepackage{bm}
\usepackage{rotating}
\usepackage{amsmath}

\newcommand{\beq}{\begin{equation}}
\newcommand{\eeq}{\end{equation}}
\newcommand{\beqs}{\begin{subequations}}
\newcommand{\eeqs}{\end{subequations}}
\newcommand{\bea}{\begin{eqnarray}}
\newcommand{\eea}{\end{eqnarray}}

\newcommand{\bfg}{\begin{figure}}
\newcommand{\efg}{\end{figure}}

\begin{document}

\title{Pseudo Slice Energy Spread  \\
in Dynamics of Electron Beams Moving through Magnetic Bends}

\author{Rui Li}
\affiliation{
Thomas Jefferson National Accelerator Facility,\\
12000 Jefferson Ave., Newport News, VA 23606, USA \\
}%
\date{\today}

\begin{abstract}
In the previous canonical formulation of  beam dynamics for an electron bunch moving ultrarelativistically through magnetic bending systems, we have shown that the transverse dynamics equation for a particle in the  bunch  has a driving term  which behaves as the centrifugal force caused by the particle's initial potential energy  due to collective particle interactions within the bunch. As a result, the initial potential
energy  at the entrance of a bending system, which we call pseudo (kinetic) energy, is indistinguishable from the usual  kinetic energy offset from the design energy in its perturbation to particle  optics through  dispersion and momentum
compaction.
In this paper, in identifying  this centrifugal force on particles  as the remnant of the CSR cancellation effect in transverse particle dynamics, we show  how  the 
dynamics  equation in terms of  the canonical momentum for beam motion on a curved orbit is related to the Panofsky-Wenzel theorem for wakefields for beam motion on a straight path. It is shown
that the effect of pseudo energy spread can be measurable only for a high-peak-current bunch when the  pseudo  slice energy spread is appreciable compared to the  slice kinetic energy spread.
The implication of the  pseudo slice energy spread for bunch dynamics in magnetic  bends is discussed.

\end{abstract}

\maketitle

\section{\label{sec:introduction}Introduction}

The design of 4th generation light sources and future  linear colliders demand high brightness electron beams with high peak current. In these designs the slice energy spread  is an important quantity that characterizes the longitudinal phase space property of the beams. For example, the high peak current is usually achieved by compressing the electron bunches using  magnetic chicanes,  with the maximum peak current after compression  being  determined by the initial longitudinal phase space distribution, including both  the slice energy spread  of the bunch and the nonlinear variation of average slice  energy offset along the bunch. The slice energy spread also plays a critical role in Landau damping of the microbunching instability, which is parasitically induced by collective interaction such as the longitudinal space charge (LSC) \cite{lsc-uBI} or coherent synchrotron radiation (CSR)  \cite{csr-uBI1,csr-uBI2} as a high brightness electron beam  is transported through beamlines including straight section or linac followed by one or more magnetic chicanes.

Accurate measurement of slice energy spread poses significant challenge to modern diagnostics because of the limited achievable resolution.  Direct measurements are often carried out using transverse deflecting RF structure (TDS) followed by a spectrometer \cite{tds}. Indirect methods, which use the slice energy spread as a fitting parameter,  include  measurement of the average  power  spectrum  of COTR generated by a microbunched beam \cite{fel08} or measurement of the coherent harmonic radiation generated by a seeded FEL \cite{chg}.  The consistency of  results for slice energy spread measured from  different schemes and their comparison with simulation results are crucial topics for investigation.

In this paper, we present a mechanism in 2D CSR interaction which can play the same role as the slice kinetic energy spread does during beam transport in bending systems. From analyses of the dynamics equation in terms of canonical momentum on a curved orbit, it was shown earlier \cite{hbb, epac02, jlabtn} that when a bunch moving ultrarelativistically  through a  bending system, there exists cancellation between the Talman's force and the integrated effect of noninertial space charge force in their joint impact on the bunch emittance growth. After this cancellation  there is a remnant centrifugal force term left which is related to the particle's initial potential energy at the entrance of the bending system. Consequently, the initial slice potential energy spread of the bunch, which we call pseudo slice energy spread, is indistinguishable from the usual slice kinetic energy spread in its perturbation  to the transverse particle optics via both dispersion and momentum compaction. The possible effect of the potential energy on beam optics in dispersive regions has been pointed out in our earlier studies \cite{hbb, epac02, jlabtn, app, dis, lir08}.  The focus of this paper is to show the origin of the pseudo slice energy spread in beam transport through bends, and to present quantitative estimates for some simple examples. 

The paper is organized as follows.  In Sec.\,II we review  previous analyses of the cancellation effect, and explain that the effect of pseudo slice energy spread can only be revealed  when both  
of the transverse and longitudinal CSR forces  are taken into account. Earlier it was pointed out \cite{pw} that  the Panofsky-Wenzel theorem is a direct result of the dynamics equation of a particle in terms of its canonical momentum on a straight path. Here it is shown that our analysis of the cancellation effect  is a straightforward generalization of this previous theory to dynamics on a curved orbit. It is shown in Sec.\,III  that the effect of pseudo slice energy spread is observable only for a  high-peak-current bunch being transported through magnetic bends
when the pseudo slice energy spread is appreciable compared to the slice kinetic energy spread. 
Quantitative examples are given for simplified examples.   The implication of pseudo slice energy spread for the studies of bunch dynamics in bends, including microbunching instability, is discussed   in Sec.\,IV .

\section{\label{sec:eqCM} Transverse Dynamics on a Curved Orbit with 2D CSR Effects }

For collective interactions of a bunch moving on a straight section, the relationship between the longitudinal and transverse wakefields on particles  is given by the  Panofsky-Wenzel theorem.
Even though this  relation is often not apparent as seen directly from  expressions of the  transverse and longitudinal Lorentz forces    (${\bf F}_{\perp}$ and $F_s$) in terms of  the  {\bf E} and {\bf B}  fields, it can be readily understood from the dynamics equation in terms of the canonical momentum. In this section, we show that our previous formulation \cite{jlabtn}  of the 2D CSR effect is a generalization of the analysis of canonical momentum for motion on straight path to that on a curved orbit. The discussion of the cancellation between the effects of centrifugal space charge force and the non-inertial (or sometimes non-conventional) space charge force on  transverse dynamics of a particle is briefly  reviewed, and the centrifugal force associated with the  initial potential energy of the particle is recognized as the remnant of this cancellation.

\subsection{The Panofsky-Wenzel Theorem and the Canonical Momentum Analysis for Bunch  Motion on a Straight Path }

We first summarize the discussion on the Panofsky-Wenzel theorem given by G. Stupakov \cite{pw}. For an electron bunch  moving on a straight path, let  {\bf E}  and  {\bf B} be either the external RF fields or the collective EM fields due to bunch self-interaction. The longitudinal and transverse wake potentials on a particle in the bunch are related to the change of momentum during the passage of the bunch through the straight section:
\bea
&& W_{l}({\bf r},s) =-\frac{c}{q}\Delta p_s =-\frac{c}{q} \int dt \, E_{z}|_{z=ct-s},  \label{Wl} \\
&& {\bf W}_{\perp}({\bf r},s) =-\frac{c}{q}\Delta {\bf  p}_{\perp} =\frac{c}{q} \int dt \, [{\bf E}_{\perp}+\hat{z}\times {\bf B}]|_{z=ct-s}. \label{Wt}
\eea
The  theorem states
\beq
\frac{\partial {\bf W}_{\perp}}{\partial{s}}={\bm \nabla}_{{\perp}} W_{l}.
\label{PW}
\eeq
This relation exists because fundamentally the motion is governed by the  principle of extremal action. With $(\Phi,{\bf A})$ denoting the 4-potentials on the relativistic charged particle and ${\bm \beta}={\bf v}/c$, the Lagrangian for the particle is
\beq
L=L_0+ L_{\mbox{\scriptsize int}},
\eeq
where the free particle Lagrangian and interaction Lagrangian are respectively
\beq
L_{0}=-mc^2 \sqrt{1-\frac{v^2}{c^2}},   \hspace{0.5 in}  L_{\mbox{\scriptsize int}}=-e(\Phi-{\bm \beta}\cdot {\bf A} ).
\eeq
For canonical momentum ${\bf P}={\bf p}+e{\bf A}/c$, the Euler-Lagrange equation yields   (with $ {\bm \nabla}$  only operating on $\Phi$ and ${\bf A}$)
\beq
\dot{{\bf P}}\equiv d{\bf P}/dt={\bm \nabla}  L_{\mbox{\scriptsize int}}.
\label{dPt}
\eeq
The time integral of the above equation,  with $t_i$ and $t_f$ denoting the time before and after the passage of  the section, gives
\bea
\Delta {\bf p}_{\perp}+e\Delta {\bf A}_{\perp}/c &=&  \int_{t_i}^{t_f} dt \, {\bm \nabla}_{\perp}  L_{\mbox{\scriptsize int}}, \\
\Delta p_{s}+e\Delta A_{s}/c &=&  \int_{t_i}^{t_f} dt \,  \nabla_{s}  L_{\mbox{\scriptsize int}} ,
\eea
for  $\Delta {\bf p}_{\perp}= {\bf p}_{\perp}(t_f)-{\bf p}_{\perp}(t_i)$ and similarly for $\Delta p_{s}$, $\Delta {\bf A}_{\perp}$ and $\Delta A_{s} $.
Recall  that the impedance and wake function are calculated assuming the bunch remains rigid during its transport through the section of interest. By  further
assuming that  the boundary condition around the bunch  is the same at  $t_i$ and $t_f$,  one has $\Delta A_{\perp}=\Delta A_{s}=0$, and thus
\beq
\frac{ \partial  (\Delta {\bf p}_{\perp})}{\partial s} ={\bm \nabla}_{\perp} (\Delta p_s),
\eeq 
which yields the Panofsky-Wenzel theorem (Eq.\,(\ref{PW})) with the help of  Eqs.\,(\ref{Wl}) and (\ref{Wt}). Note that  numerical verification of Eq.\,(\ref{PW})  requires the knowledge of  both  longitudinal and transverse Lorentz forces on particles.

\subsection{Generalization  to Motion on a Curved Orbit}

We now give a brief review of the cancellation effect in transverse dynamics under CSR induced perturbation \cite{epac02, jlabtn}. The origin of pseudo  energy spread is revealed and explained in this review.

For  a beam moving relativistically (in free space) on a circular orbit with design radius R, the particle dynamics in the bending plane can be expressed in cylindrical coordinates with respect to the center of the circular orbit: ${\bf r}=r{\bf e}_r$ and ${\bf v}=\dot{r} {\bf e}_r + r\dot{\theta} {\bf e}_s$. The Lagrangian for a particle is $L=L_0+ L_{\mbox{\scriptsize int}}$, with
\beq
L_0=-mc^2 \sqrt{1-\frac{r^2\dot{\theta}^2+\dot{r}^2}{c^2}},\hspace{0.5in}  L_{\mbox{\scriptsize int}}=-e\left (\Phi-\frac{\dot{r}}{c}A_r -\frac{r\dot{\theta}}{c}A_s \right).
\eeq
 We have shown in Ref.\,\cite{jlabtn} the equivalence of Euler-Lagrange equations in terms of curvilinear coordinates and the dynamics equations  obtained by projecting Eq.~(\ref{dPt}) to the local radial-azimuthal bases. Here we outline the latter approach. Denoting $s=R\theta$ as the path length on the circular orbit, we have
 $d{\bf e}_r/ds={\bf e}_s/R$ and   $d{\bf e}_s/ds=-{\bf e}_r/R$, or $\dot{{\bf e}_r}=\dot{\theta}{\bf e}_s$  and $\dot{{\bf e}_s}=-\dot{\theta}{\bf e}_r$. 
With the components of canonical momentum of a particle being
\beq
P_r= p_r + eA_r/c, \hspace{0.5in} P_s=p_s +eA_s/c,
\eeq
for  $p_r=\gamma m\dot{r}$ and  $p_s=\gamma m r\dot{\theta}$ representing the kinetic momentum components,
one can directly generalize  Eq.~(\ref{dPt}) for motion on a straight path to motion on a circular orbit by using 
\beq
\frac{d ({\bf P}\cdot {\bf e}_r)}{dt}= \frac{d {\bf P}}{dt}\cdot {\bf e}_r+{\bf P} \cdot \frac{d {\bf e}_r}{dt},
\eeq
which yields
\beq
\frac{dP_r}{dt}-v_s \frac{P_s}{r} = \frac{\hat{\partial}  L_{\mbox{\scriptsize int}}}{\partial r}.
 \label{dPr} 
\eeq
Here  $\hat{\partial}$ denotes the differential operator  acting only on  $\Phi$ and ${\bf A}$, e.g.,
\beq
\hat{\partial}_r L_{\mbox{\scriptsize int}}=
-e(\partial_r \Phi- \beta_r  \partial_r A_r-\beta_s  \partial_r A_s).
\eeq
Note in  Eq.~(\ref{dPr}), $(-v_s P_s/r ){\bf e}_r=P_s \dot{{\bf e}}_s$ is the geometrical term accounting  for the rotation of unit vector ${\bf e}_s$ tangent to the design orbit. 
Similar to Eqs.~(\ref{dPr}), with the Hamiltonian (canonical energy) 
\beq
H=c\sqrt{({\bf P}-e{\bf A}/c)^2+m^2c^2}+e\Phi
\eeq
and kinetic energy $\mathcal{E}_k=\gamma mc^2$, the energy equation $dH/dt=\partial H/\partial t$ yields
\beq
\frac{d(\mathcal{E}_k+e\Phi)}{dt}=-\frac{\hat{\partial} L_{\mbox{\scriptsize int}}}{\partial t},
\label{Eft}
\eeq
which is consistent with ${\bf E}=-\nabla \Phi -\partial {\bf A}/c\partial t$ and thus
\beq
\frac{d\mathcal{E}_k}{dt}={\bf v}\cdot e{\bf E}=-e \frac{d \Phi}{dt}+ e \frac{\hat{\partial}(\Phi -{\bm \beta}\cdot {\bf A})}{\partial t}.
\label{dEkdt}
\eeq

The potentials $(\Phi, {\bf A})$  in the above equations consist of contributions from both  the external fields and the fields for bunch collective interactions:
\beq
\Phi=\Phi^{\mbox{\scriptsize ext}}+\Phi^{\mbox{\scriptsize col}}, \hspace{0.2in} {\bf A}={\bf A}^{\mbox{\scriptsize ext}}+{\bf A}^{\mbox{\scriptsize col}}.
\eeq
The  external magnetic field associated with the  design  energy $\mathcal{E}_0=\gamma_0 mc^2$ and the design radius $R$ is
\beq
{\bf B}^{\mbox{\scriptsize ext}}=\frac{p_0 c}{eR}({\bf e}_s \times {\bf e}_r)={\bm \nabla}\times {\bf A}^{\mbox{\scriptsize ext}},
\vspace{0.1in}
\eeq
with $p_0=\gamma_0 \beta_0 mc$ for $\beta_0=(1-\gamma_0^{-2})^{1/2}$. Here ${\bf B}^{\mbox{\scriptsize ext}}$ can be obtained from
\beq
\Phi^{\mbox{\scriptsize ext}}=0,
\hspace{0.2in}
{\bf A}^{\mbox{\scriptsize ext}} =-\frac{p_0 c}{e} \frac{r}{2R} {\bf e}_s.
\eeq
This corresponds to the external Lorentz force (including the  centripetal force) on particles
\beq
{\bf F}^{\mbox{\scriptsize ext}}=\frac{e}{c}{\bf v} \times {\bf B}^{\mbox{\scriptsize ext}}= -\frac{v_s p_0}{R}{\bf e}_r + \frac{v_r p_0}{R}{\bf e}_s.
\eeq
Unlike the external fields,  the  collective EM fields   in the vicinity of the bunch comove with the bunch.
This motivate us to define the part of  canonical momentum and interaction Lagrangian related to the collective interaction potentials
$(\Phi^{\mbox{\scriptsize col}}, {\bf A}^{\mbox{\scriptsize col}})$ as following: 
\beq
P_s^{\mbox{\scriptsize col}}=p_s+e A_s^{\mbox{\scriptsize col}}/c,
\label{Ps}
\hspace{.2in} 
P_r^{\mbox{\scriptsize col}}=p_r+e A_r^{\mbox{\scriptsize col}}/c
\eeq
and
\beq
L_{\mbox{\scriptsize int}}^{\mbox{\scriptsize col}}=-e(\Phi^{\mbox{\scriptsize col}}- \beta_r  A_r^{\mbox{\scriptsize col}}-\beta_s  A_s^{\mbox{\scriptsize col}}).
\label{lint}
\eeq
Here we use the retarded potentials for the collective interaction potentials
\beq
\Phi^{\mbox{\scriptsize col}}({\bf x},t)=\int \frac{\rho({\bf x'}, t-|{\bf x}-{\bf x'}|/c)}{|{\bf x}-{\bf x'}|} d{\bf x'},
\hspace{0.2in}
{\bf A}^{\mbox{\scriptsize col}}({\bf x},t)=\frac{1}{c} \int \frac{{\bf J}({\bf x'}, t-|{\bf x}-{\bf x'}|/c)}{|{\bf x}-{\bf x'}|} d{\bf x'}.
\label{ret}
\eeq
This choice of gauge  is natural for the exhibition of the cancellation effect \cite{jlabtn} since the local contributions $(\mbox{when} \,\, {\bf x'}\rightarrow {\bf x})$ to $\Phi^{\mbox{\scriptsize col}}$ and  ${\bf A}^{\mbox{\scriptsize col}}$ dominate and consequently $A_s \sim \beta_s \Phi$ at $v_s \simeq c$.

With the external potentials separated from the collective ones, the dynamics equation in  Eqs.~(\ref{dPr}) and ~(\ref{dEkdt}) can be organized in two different ways. The first  is  obtained by rewriting Eq.~(\ref{dPr}) as
\beq
\frac{d(\gamma m \dot{r})}{dt}=\frac{v_s P_s^{\mbox{\scriptsize col}} (t)}{r}-\frac{v_s p_0}{R}+F_r^{\mbox{\scriptsize eff}}
\label{dprt0}
\eeq
for
\beq
F_r^{\mbox{\scriptsize eff}}=-\frac{e}{c}\frac{dA_r^{\mbox{\scriptsize col}}}{dt} +\frac{\hat{\partial} L_{\mbox{\scriptsize int}}^{\mbox{\scriptsize col}}}{\partial r}.
\eeq
Here the transverse dynamics equation, Eq.~(\ref{dprt0}),  indicates that the transverse kinetic momentum is  changed by the {\it total} centrifugal force  
$v_s  P_s^{\mbox{\scriptsize col}}/r$, the external radial force $(-v_s p_0/R)$ and the effective
transverse CSR force $F_r^{\mbox{\scriptsize eff}}$. This formula emphasizes on the {\it total} centrifugal force experienced by the charged particle as the geometrical effect associated with the {\it  canonical momentum} $ P_s^{\mbox{\scriptsize col}}$, in which  the usual centrifugal force $v_s p_s/r$ 
related to the  kinetic momentum   works together with the centrifugal space charge force $F^{\mbox{\scriptsize CSCF}}$
\beq
\frac{v_s P_s^{\mbox{\scriptsize col}} (t)}{r}=\frac{v_s p_s (t)}{r}+F^{\mbox{\scriptsize CSCF}},
\label{Psr}
\hspace{0.2in}
F^{\mbox{\scriptsize CSCF}}=\beta_s \frac{eA^{\mbox{\scriptsize col}} _{s}}{r}.
\eeq
After getting Eq.~(\ref{dprt0}) for the transverse dynamics from Eq.~(\ref{dPr}), we can further obtain equation for energy from Eq.~(\ref{dEkdt}):
\beq
\mathcal{E}_{k}(t)+e\Phi^{\mbox{\scriptsize col}}(t)=\mathcal{E}_{k}(0)+e\Phi^{\mbox{\scriptsize col}}(0) + \int_0^t F_v^{\mbox{\scriptsize eff}}(t') \,c dt',
\label{dEf}
\eeq
which implies the total canonical energy is changed by the effective longitudinal CSR force
\beq
F_v^{\mbox{\scriptsize eff}}=\frac{e}{c} \left(\frac{\partial \Phi^{\mbox{\scriptsize col}}}{\partial t}- {\bm \beta} \cdot \frac{\partial {\bf A}^{\mbox{\scriptsize col}}}{\partial t}\right).
\label{Fv}
\eeq
Here $\Phi^{\mbox{\scriptsize col}}(t)$ in Eq.~(\ref{dEf})  is the brief expression of $\Phi^{\mbox{\scriptsize col}}({ \bf x}(t),t)$ with  ${ \bf x}(t)$  representing the trajectory of the particle.
With $p_s=\beta_s \mathcal{E}_k/c $ in Eq.~(\ref{Psr}) expressed in terms of
$\mathcal{E}_{k}(t)$ in Eq.~(\ref{dEf}), one can   rewrite  Eq.\,(\ref{dprt0}) as
\beq
\frac{d(\gamma m \dot{r})}{dt}+ v_s \left(\frac{p_0}{R}-\frac{p_0 }{r}\right) =  \frac{\beta_s [P_s^{\mbox{\scriptsize col}} (t)-p_0]c}{r} +F_r^{\mbox{\scriptsize eff}} 
\label{dpr}
\eeq
with
\beq
[P_s^{\mbox{\scriptsize col}}(t)-p_0]c = \beta_s \Delta \mathcal{E}^{\mbox{\scriptsize tot}}+e   [A_s^{\mbox{\scriptsize col}}(t)- \beta_s \Phi^{\mbox{\scriptsize col}}(t)] +\beta_s \int_0^t F_v^{\mbox{\scriptsize eff}}(t') \,dt'
\label{pstot}
\eeq
for $\Delta \mathcal{E}^{\mbox{\scriptsize tot}}$ denoting the total canonical energy deviation from the design energy  ($v_s \simeq c$)
\beq
\Delta \mathcal{E}^{\mbox{\scriptsize tot}}=  \Delta \mathcal{E}_k (0) + e\Phi^{\mbox{\scriptsize col}} (0), 
\hspace{0.2in}
\Delta \mathcal{E}_k(0)  \approx \mathcal{E}_k(0) -\mathcal{E}_0.
\eeq
Note that the contribution from local interaction  (when ${\bf x'} \rightarrow {\bf x}$)  can cause  both $\Phi^{\mbox{\scriptsize col}}(t)$ and $A_s^{\mbox{\scriptsize col}}(t)$ 
to have logarithmic-like dependence on the particles'  transverse deviation from the bunch center, yet such sensitivity is largely canceled in their combined effect  such as in the term  $e[A_s^{\mbox{\scriptsize col}}(t)- \beta_s \Phi^{\mbox{\scriptsize col}}(t)]$ of
Eq.~(\ref{pstot}) and in the derivatives of $L_{\mbox{\scriptsize int}}^{\mbox{\scriptsize col}}$ of  Eq.\,(\ref{lint}).
With $e   [A_s^{\mbox{\scriptsize col}}(t)- \beta_s \Phi^{\mbox{\scriptsize col}}(t)] $ negligible in Eq.~(\ref{pstot}) \cite{epac02, jlabtn, dis}, we have
\beq
[P_s^{\mbox{\scriptsize col}}(t)-p_0]c \approx  \beta_s \Delta \mathcal{E}^{\mbox{\scriptsize tot}}+\beta_s \int_0^t F_v^{\mbox{\scriptsize eff}}(t') \,c dt'.
\label{Ps0}
\eeq
 It is important to note that after applying Eq.~(\ref{Ps0})  to  Eq.~(\ref{dpr}),
the sensitive dependence of the driving terms on the transverse coordinates of particles only shows up in the $e\Phi^{\mbox{\scriptsize col}}(0)/r$ term---the centrifugal force
related to the initial potential energy (or pseudo kinetic energy) which is the main focus  of this paper.

The second way to organize Eqs.\,(\ref{dPr}) and (\ref{dEkdt}) follows the usual approach emphasizing on the  particle dynamics governed  by Lorentz force 
${\bf F}^{\mbox{\scriptsize col}}=e({\bf E}^{\mbox{\scriptsize col}}+{\bm \beta}\times {\bf B}^{\mbox{\scriptsize col}})=F_r^{\mbox{\scriptsize col}} {\bf e}_r
+ F_s^{\mbox{\scriptsize col}} {\bf e}_s$ due to the collective interaction of a bunch of particles on a curved orbit,
with the transverse dynamics of particles described by
\beq
\frac{d(\gamma m \dot{r})}{dt}+ v_s \left(\frac{p_0}{R}-\frac{p_s }{r}\right) = F_r^{\mbox{\scriptsize col}}, 
\label{udpr}
\eeq
for \cite{derb2}
\beq
 F_r^{\mbox{\scriptsize col}}=
  F^{\mbox{\scriptsize CSCF}} +  F_r^{\mbox{\scriptsize eff}},
\label{dprt} 
\eeq
and the energy change described by
\beq 
\frac{d\mathcal{E}_k}{cdt}={\bm \beta} \cdot {\bf F}^{\mbox{\scriptsize col}}=  F^{\mbox{\scriptsize NSCF}}+F_v^{\mbox{\scriptsize eff}},
\label{dEcdt}
\eeq
with $F_v^{\mbox{\scriptsize eff}}$ from Eq.~(\ref{Fv}) and  the  non-inertial (or sometimes non-conventional) space charge force induced from potential energy change
\beq
F^{\mbox{\scriptsize NSCF}}=-e\frac{d\Phi^{\mbox{\scriptsize col}}}{c dt}.
\label{fnscf}
\eeq
This yields energy relation equivalent to Eq.~(\ref{dEf})
\beq
\mathcal{E}_k(t)=\mathcal{E}_k (0) - e [\Phi^{\mbox{\scriptsize col}}(t)-\Phi^{\mbox{\scriptsize col}}(0)]+ \int_0^t F_v^{\mbox{\scriptsize eff}}(t') \,c dt'.
\label{dEf1}
\eeq

On the left hand side  (LHS) of Eq.\,(\ref{udpr}), the term  $-v_s p_s/r$ depends on the energy variation $\mathcal{E}_k (t)$ according to  $p_s=\beta_s \mathcal{E}_k(t)/c$, 
with $\mathcal{E}_k(t)$ determined by the space charge and CSR interaction  as described by Eq.~(\ref{dEf1}).
This leads to the time dependent and transverse-coordinate sensitive term $e\beta_s^2 \Phi^{\mbox{\scriptsize col}}(t)/r$  on the LHS of Eq.~(\ref{udpr}), which is
  largely canceled by a similar term $F^{\mbox{\scriptsize CSCF}}$  in the {\it transverse} Lorentz force $F_r^{\mbox{\scriptsize col}}$ 
on the right hand side (RHS) of Eq.\,(\ref{udpr}). As previously mentioned, other than the term related to the  initial potential energy, $e\beta_s^2 \Phi^{\mbox{\scriptsize col}}(0)/r$,  the remaining terms have weak dependence on particle transverse coordinates.
This cancellation reflects the close interplay of the potential energy variation and the transverse CSR force in their joint effects on  the transverse particle dynamics in a magnetic dipole.
Just as  the Panofsky-Wenzel theorem, this close interplay may not be apparent from the point of view of Lorentz forces. However, it can be readily perceived from   Eqs.~(\ref{dprt0})  and (\ref{Ps0}): the former shows that the total centrifugal force is a geometrical
effect of the  longitudinal canonical momentum as a whole, and the latter shows that regardless of the redistribution of
the kinetic and potential energy of a particle as it is transported through the  bending system, with each of them sensitive to the transverse coordinates of the particle, the
{\it change} of the longitudinal canonical momentum as a whole over time is the integral of the
effective longitudinal force which depends mainly on the longitudinal coordinate of the particle in the bunch.

Under the assumptions
\beq
\left\{ \delta=\frac{\mathcal{E}-\mathcal{E}_0}{\mathcal{E}_0},\\ \gamma^{-1}, \\ x', \\  \frac{x}{R}, \\ \frac{ I_p}{\gamma I_A},  \\ \frac{x/R}{(l/R)^{1/3}}\right\} \ll 1,
\label{small}
\eeq
with  $l$ the characteristic modulation length in the bunch and $I_p$  the bunch peak current,
we see from Eqs.\,(\ref{dprt0}) or (\ref{udpr}) how the usual linear horizontal optics is  perturbed by the  CSR effect 
\beq
\frac{d^2 x}{ds^2}+ \frac{x}{R^2}\simeq \frac{\delta_k(0)}{R}+G^{\mbox{\scriptsize col}},
\label{dynx}
\eeq
with $s=ct$ the path length of a particle,  $\delta_k (0)=(\mathcal{E}_k-\mathcal{E}_{k0})/\mathcal{E}_0$ the initial relative  kinetic energy offset, and $G^{\mbox{\scriptsize col}}$   representing the joint effect of the  horizontal and longitudinal CSR forces 
\beq
G^{\mbox{\scriptsize col}}=G_{\phi 0}+ G_{\mbox{\scriptsize res}}+G_v+G_x.
\eeq
Here $G_{\phi 0}$ represents the effect of  initial  potential energy offset on transverse dynamics
\beq
G_{\phi 0}=\frac{\delta_{\phi}(0)}{R},
\hspace{0.2in}\mbox{with}\hspace{0.2in}
\delta_{\phi}(0) \simeq  \frac{e \Phi^{\mbox{\scriptsize col}} (0)}{\mathcal{E}_{0}},
\label{Gf0}
\eeq
$G_v$ and $G_x$ are related to the effective forces
\beq
G_v=\frac{ \int_0^t F_v^{\mbox{\scriptsize eff}}(t') \,c dt'}{R  \mathcal{E}_{0}},
\hspace{0.4in}
G_x=\frac{F_x^{\mbox{\scriptsize eff}}}{\mathcal{E}_0},
\eeq
and $G_{\mbox{\scriptsize res}}$  represents the residual of the cancellation, which is of the second order of small quantities in Eq.~(\ref{small}),
\beq
G_{\mbox{\scriptsize res}}=\frac{e[A_s^{\mbox{\scriptsize col}}(t)-\Phi^{\mbox{\scriptsize col}}(t)]}{R\mathcal{E}_0}.
\eeq 
Single particle optics resumes when $G^{\mbox{\scriptsize col}}=0$ in Eq.~(\ref{dynx}).

For a beam being transported  through a bending system, Eq.~(\ref{dynx}) can be rewritten as
\beq
\frac{d^2 x}{ ds^2}+ \frac{x}{R^2(s)} \simeq  \frac{\delta_{k0}+\delta_{\phi 0}}{R(s)}+\hat{G}^{\mbox{\scriptsize col}}(s),
\label{dynx1}
\eeq
with $R(s)\rightarrow\infty$ for straight sections,   $\delta_{k0}=\delta_{k}(0)$, $\delta_{\phi 0}=\delta_{\phi}(0)$, and 
\beq
\hat{G}^{\mbox{\scriptsize col}}= G_{\mbox{\scriptsize res}}+G_v+G_x \simeq G_v+G_x
\eeq
standing for the CSR  perturbation terms  related to the effective forces. It has been shown \cite{derb2, hbb} that  the  effective forces  are insensitive to the particle transverse coordinates (more details will be discussed in Sec.\,\ref{sec:dscl}). 
Combining with $dz/ds=-x/R(s)$, one then finds that
the {\it  initial} kinetic energy offset and potential energy of a particle at the entrance of a bending system always work together for  their dispersive impact on  single partical optics, namely,
\beq
\left\{\begin{array}{l}
x = R_{11} x_0 + R_{12} x'_0 + R_{16} (\delta_{k 0}+\delta_{\phi 0}) +\Delta x_c  \\
x'= R_{21} x_0 + R_{22} x'_0 + R_{26} (\delta_{k 0}+\delta_{\phi 0}) +\Delta x'_c  \\
z =  z_0+ R_{51} x_0 + R_{52} x'_0 + R_{56} (\delta_{k 0}+\delta_{\phi 0}) +\Delta z_c  \\ 
(\delta_k + \delta_{\phi} )  = (\delta_{k 0}+ \delta_{\phi 0}) + \Delta \delta_{kc}
\end{array}\right.
\label{xc}
\eeq
Here  $R_{ij}$ ($i=1$ to 6, $j=1$ to 6)  are the elements of the usual transport matrix $R(0\rightarrow s)$,   $(\Delta x_c, \Delta x'_c , \Delta z_c)$    come from effects of $\hat{G}^{\mbox{\scriptsize col}}$, and $\Delta \delta_{kc} $ is resulted from $\int F_v^{\mbox{\scriptsize eff}}(t')cdt' $ in Eq.\,(\ref{dEf}) or Eq.\,(\ref{dEf1}).
Here $G_{\phi 0}$  is related to the {\it initial} potential energy of the particle, and the signifi-cance of its impact
on beam dynamics depends on the comparison of the initial potential energy spread with the initial kinetic energys spread.
On the other hand, $\hat{G}^{\mbox{\scriptsize col}}$ is related to the correlated perturbation taking place {\it during} the beam transport through the bending system, in which  $F_v^{\mbox{\scriptsize eff}}$ often plays the dominant role while the impact of $F_r^{\mbox{\scriptsize eff}}$ depends on its comparison with that of  $F_v^{\mbox{\scriptsize eff}}$.
More  detailed analysis using Frenet frame coordinates can be found in Ref.\,\cite{lir08, acrt}.  The reason we name $\delta_{\phi 0}$ the relative pseudo (kinetic) energy,  or name $e\Phi^{\mbox{\scriptsize col}}(0)$ the pseudo (kinetic) energy, is that without detailed analysis of the interplay of 2D CSR forces,
one tends to attribute the measured result of the slice energy spread  in dispersive regions as caused by the slice kinetic energy spread alone, and therefore miss the fact that part of the measured result could be contributed from the potential energy as described in Eq.\,(\ref{xc}).

It is necessary to point out that as an electron bunch moves through a bending system, its total canonical energy $\mathcal{E}=
\mathcal{E}_k +e\Phi^{\mbox{\scriptsize col}}$ varies from  the entrance of one dipole magnet to the entrance of another dipole following Eq.\,(\ref{dEf}), as the result of the effective longitudinal CSR force
$F_v^{\mbox{\scriptsize eff}}$ along the beam line. However, we can still use $(\delta_{k0}+\delta_{\phi 0})$ at the entrance of the whole bending system to summarize
the logarithmic-like dependence, since $F_v^{\mbox{\scriptsize eff}}$ to the first order is free from transverse dependence \cite{roll} and its effect is included in 
$(\Delta x_c, \Delta x'_c , \Delta z_c,\Delta \delta_{kc} )$.

\subsection{ Survey of Previous Studies and Role of $G_{\phi 0}$  as the  Remnant of  Cancellation}

The cancellation effect in CSR has been a long-standing controversial topic  and the history of  debates was reviewed earlier \cite{jlabtn, Talman2}. To identify the role of $G_{\phi 0}$ in these debates, here we survey 
some of the previous studies.  In particular, we focus on the two major parts of the cancellation, i.e., 
the centrifugal space charge force (see Eq.\,(\ref{Psr}))
\beq
F^{\mbox{\scriptsize CSCF}}=\beta_s\frac{eA_{s}^{\mbox{\scriptsize col}}}{r}
\label{Fcscf}
\eeq
and the integrated effect of the noninertial (or sometimes non-conventional) space charge force (see Eq.\,(\ref{fnscf}))
\beq
\int_{0}^{t} F^{\mbox{\scriptsize NSCF}}(t') \,cdt'=-e(\Phi^{\mbox{\scriptsize col}}(t)-\Phi^{\mbox{\scriptsize col}}(0)).
\label{Fnscf}
\eeq
It can be shown that in deriving  Eq.~(\ref{dynx})  from Eq.~(\ref{udpr}),  the terms in $G^{\mbox{\scriptsize col}}$ sensitive  to the transverse coordinates  of particles can be summarized to the first order as
\beq
\frac{1}{\mathcal{E}_0} \left(F^{\mbox{\scriptsize CSCF}}+\frac{\beta_s \int_{0}^{t} F^{\mbox{\scriptsize NSCF}}(t')cdt' }{R}\right)  
\simeq \frac{e(A_s^{\mbox{\scriptsize col}} (t)-\Phi^{\mbox{\scriptsize col}}(t))/\mathcal{E}_0}{R}+\frac{e\Phi^{\mbox{\scriptsize col}}(0)/\mathcal{E}_0}{R} \simeq G_{\phi 0},
\label{cc}
\eeq
with $G_{\phi 0}$ emerging as the net remnant of the cancellation between $F^{\mbox{\scriptsize CSCF}}$ and the integrated effect of $F^{\mbox{\scriptsize NSCF}}$ on their impacts to  the horizontal beam optics.

The existence of the centrifugal space charge force was first pointed out by Talman \cite{Talman1} in  his pioneer study of the space charge interaction for  beams moving on the circular orbit in a storage ring.
It was found that unlike the usual space charge force on a straight path, the transverse Lorentz force, expressed  in terms of the Lienard-Wiechert fields, has a term with logarithmic-like dependence on particle's transverse coordinates. This term  could lead to horizontal tune shift and equivalent chromaticity effect, and consequently the appearance of nonlinear resonances.
A following study  by Lee  showed \cite{lee} that for a coasting beam in a storage ring with $\beta\simeq 1$, the harmful impact of the Talman's force on particle transverse dynamics  is canceled by the effect of kinetic energy change. This is because  in Eq.\,(\ref{dEf}) we have $F_v^{\mbox{\scriptsize eff}}\simeq 0$ for the coasting beam.  Hence for $E_e$ and $r_e$ being the equilibrium energy and radius, one has
\beq
\mathcal{E}_k(t)+e\Phi^{\mbox{\scriptsize col}}(r(t)) =\mathcal{E}_{ke} + e\Phi^{\mbox{\scriptsize col}} (r_e),  \hspace{0.2in} \mbox{ or} \hspace{0.2in}  \mathcal{E}(t)=- e\Phi ^{\mbox{\scriptsize col}}(r(t)) + \mbox{ constant}. 
\label{lee}
\eeq
It was shown that the  corresponding radial  force $-e\Phi^{\mbox{\scriptsize col}}(r(t)) /r$  largely cancels with the centrifugal space charge force $eA^{\mbox{\scriptsize col}}(t)/r$, with the  transverse sensitivity of the remaining terms negligible compared to the two  leading terms involved in the cancellation. More discussion on this problem can be found in Ref.\,\cite{app}.

With the   increasing demand for  linac drivers of  FEL to provide electron beams with  low emittance and high peak current, interests  in the CSR effect shifted from coasting beams in  storage rings to bunched beams in beamlines including general magnetic bending systems. In an analysis of the transverse CSR force on electron bunches, Derbenev concluded \cite{derb2}  that the impact of Talman's force on transverse dynamics always cancels with the impact of kinetic energy change due to the change of potential energy. Meanwhile,
Carlsten studied the interaction of an off-axis particle interacting with an electron bunch on a design circular orbit, and found that in addition to the transverse Talman's force and the usual longitudinal CSR force, there exists another term of longitudinal CSR force \cite{nscf}. This term is named non-inertial space charge force, or $F^{\mbox{\scriptsize NSCF}}$, which represents  the space-charge curvature effect and causes modification of particle's energy with little total loss by radiation. It was noted that both the two longitudinal CSR forces, the new $F^{\mbox{\scriptsize NSCF}}$ and the usual one, will cause redistribution of particle's kinetic energy within an achromatic bend system and can cause emittance growth. It was later recognized  \cite{hbb} that  $F^{\mbox{\scriptsize NSCF}}$ in Carlsten's example (which is the combination of the 2nd and 3rd terms in Eq.\,(9) of Ref.\,\cite{nscf}), to the first order, is $-ed\Phi^{\mbox{\scriptsize col}}/cdt$ (see Eq.\,(48) of Ref.\,\cite{hbb}). This is the origin of our definition of  $F^{\mbox{\scriptsize NSCF}}$ in  Eq.\,(\ref{fnscf}), even though in general $d\Phi^{\mbox{\scriptsize col}}/dt$ can be induced by many more ways of particle interaction than that discussed in the original  example \cite{nscf} of CSR interaction for  off-axis particles . With this identification of $F^{\mbox{\scriptsize NSCF}}$, one finds  \cite{hbb}  that its integrated effect  cancels with the effect of Talman's force, as summarized in Eq.\,(\ref{cc}), in their joint impact  on the transverse emittance growth in an achromatic bending system. 

In a following study on the effect of space charge interaction in a bunch compression chicane, it was pointed out by Bane and Chao \cite{bnchao}  that because of the drastic beam size convergence during the drift between the last two bends, the longitudinal space charge force is no longer proportional to $\gamma^{-2}$.  This will cause changes in particle kinetic energy, which will further lead to emittance growth  in the last bend of the chicane. This is again the effect of $F^{\mbox{\scriptsize NSCF}}$ in Eq.~(\ref{Fnscf}).  However, here it   is more appropriate to  call $F^{\mbox{\scriptsize NSCF}}$  the {\it  non-conventional space charge force} because it is  originated from the Coulomb interaction on straight path, as oppose to the non-inertial space charge force originated from the
radiative part of Lienard-Wiechert field \cite{nscf} on a curved orbit. It was pointed out  later   \cite{app}   that for the integrated kinetic energy change $\Delta \mathcal{E}_k=-e(\Phi^{\mbox{\scriptsize col}}(t)-\Phi^{\mbox{\scriptsize col}}(0)) + \int F_v^{\mbox{\scriptsize eff}}(t')cdt'$, the contribution of potential energy to the radial force, $-e\Phi^{\mbox{\scriptsize col}}(t)/R$, on the particle in the last bend is canceled by the transverse Talman's force $eA_s^{\mbox{\scriptsize col}}(t)/R$ on the particle. 

Finally during the analysis of the transverse CSR force in terms of the Lienard-Wiechert fields for a bunched beam, Geloni and others found \cite{mis} that instead of the commonly accepted feature of CSR interaction as a tail-head (overtaking) interaction,
there is a head-tail part in the transverse CSR force arising from interaction on a test particle generated by particles
ahead of it. Their study shows that the head-tail part is originated from the radiative part of the Lienard-Wiechert fields, and for a uniform bunch on a design orbit, it could be several times larger than the tail-head part of the transverse Lorentz force.  It is also noted that $G_{\phi 0}=e\Phi^{\mbox{\scriptsize col}}(0)/R$ includes the head-tail interaction and as a source of perturbation it cannot be canceled away. It was later discussed \cite{dis} that for the example in Ref.\,\cite{mis}, the head-tail part of transverse Lorentz force is exactly the head-tail part of $F^{\mbox{\scriptsize CSCF}}(t)$ in Eq.\,(\ref{Fcscf}),  and it is always canceled by the $e\Phi^{\mbox{\scriptsize col}}(t)/R$
term from the kinetic  energy change since the latter contains the same head-tail part as in  $F^{\mbox{\scriptsize CSCF}}$.

As the above survey indicates, the integrated effect of  $F^{\mbox{\scriptsize NSCF}}=-e d\Phi^{\mbox{\scriptsize col}}/cdt$, or potential energy change $-e[\Phi^{\mbox{\scriptsize col}}(t)-\Phi^{\mbox{\scriptsize col}}(0)]$,  can be  originated from either the radiative  or the Coulomb part of the Lorentz force. Here  $e d\Phi^{\mbox{\scriptsize col}}/dt$  is caused by the change of particle interaction,  and can take various forms such as (1) the non-inertial space charge force for an off-axis particle interacting with the bunch on a circular orbit \cite{nscf}, and (2)  the longitudinal space charge force for a converging beam in a chicane \cite{bnchao}, and (3) even in transient  CSR interaction as a bunch entering or exiting a magnetic dipole \cite{app}.  In all these cases, $e \Phi^{\mbox{\scriptsize col}}(t)/R$  cancels with  $eA^{\mbox{\scriptsize col}}_s(t)/R$ during the bunch transport on a circular orbit, leaving $G_{\phi 0}=e\Phi^{\mbox{\scriptsize col}}(0)/R$ as the remnant of this cancellation. Since $e A^{\mbox{\scriptsize col}}_{s}(t)/R$ is a part of radial Lorentz force, and $e[\Phi^{\mbox{\scriptsize col}}(t)-\Phi^{\mbox{\scriptsize col}}(0)]$ requires accurate integral of particle longitudinal dynamics, only in complete and fully self-consistent 2D/3D  treatment with both longitudinal and transverse CSR forces included, can the cancellation be taken care of naturally and thus the remnant $ G_{\phi 0} $ be revealed. Even though  $ G_{\phi 0} $   features the similar sensitive logarithmic-like dependence on the  transverse position of particles as does the Talman's force, and contains contributions from head-tail interaction, it is the effect of {\it  initial} potential energy of particles before entering the bending system and thus it plays the same role as the initial kinetic energy spread in optical transport through bends as expressed in Eq.\,(\ref{xc}). Hence $ G_{\phi 0} $  does not directly cause emittance growth for an achromatic bending system.

Presently 1D CSR codes are based on rigid-line bunch model for CSR force calculation \cite{borland}. In this  model  the transverse CSR force in Eq.\,(\ref{dprt}) is set to zero,  i.e.,  $ F^{\mbox{\scriptsize CSCF}}= F_r^{\mbox{\scriptsize eff}}=0$, and only longitudinal CSR forces  given by Eq.\,(\ref{dEcdt}) for a rigid-line bunch is applied.  For the  steady-state
interaction, only $F_v^{\mbox{\scriptsize eff}}$ applies since $F^{\mbox{\scriptsize NSCF}}=0$. 
The success of 1D CSR model in its simulation of CSR experiments and in achieving good agreements with emittance and bunch length measurements indicate that the assumptions used for the 1D model are valid for the parameter regimes in current experimental operations.  Notice that in the 1D model, one part of the cancellation,  $F^{\mbox{\scriptsize CSCF}}$, is ignored, and  the other part $\int_{0}^{t} F^{\mbox{\scriptsize NSCF}}(t')cdt'$ (from transient  interaction) only depends on $z$  for a line bunch. So the potential energy spread due to transverse particle coordinates is not included in this model. Since initial energy spread does not cause emittance growth for an achromatic bending system, the negligence of $G_{\phi 0}$ effects in the 1D CSR  model  does not prevent the model to give good prediction of the CSR induced emittance growth in a bunch compression chicane, as long as the effective transverse CSR force has negligible impact on transverse dynamics compared to that of the effective longitudinal CSR force.  However,  with full dynamics included, the relative pseudo energy spread  $\delta_{\phi 0}$  may appear wherever the relative  kinetic energy spread plays a role such as in the minimum bunch length after full compression or in the measurement of slice energy spread in dispersive regions, and its significance depends on its  quantitative comparison with the  initial slice  kinetic energy spread of the bunch. In the following section, the pseudo slice energy spread will be estimated for a simplified model.

\section{\label{sec:eFi}The Potential Energy  for a  Gaussian Bunch and  the Slice Total Energy Spread}

We are interested in  the quantitative estimation of the  potential energy  of particles at the entrance of a bending system,  which is $e \Phi^{\mbox{\scriptsize col}}(0)$    in  Eq.\,(\ref{Gf0}). We will calculate the retarded scalar potential in the  Lorentz gauge as a result  of our earlier discussion \cite{jlabtn}  that the cancellation effect  is most naturally exhibited in this framework.
 In general, for a bunch with finite emittance undergoing optical transport through a straight path, the calculation of the retarded potentials needs to take into account 
the history of particle dynamics for all particles in the bunch. However, since this study aims at illustrating the effect of potential energy on the optics of particles, we only limit ourselves to a simple estimation of the potential energy dependence on particle position  for a rigid 3D 
Gaussian bunch moving on a straight path. The analysis of the retarded scalar potential for a Gaussian bunch, as  detailed in Appendix A, will be applied
to find the probability distribution of particles in potential energy. This will subsequently be  used to determine the slice spread of total (canonical) energy considering  the fact that transverse optics is perturbed by the dispersive effect of both the kinetic and potential energy offset of particles  together (Eq.~(\ref{xc}) ).

\begin{figure}[h]
\includegraphics[width=7cm,height=7cm,keepaspectratio]{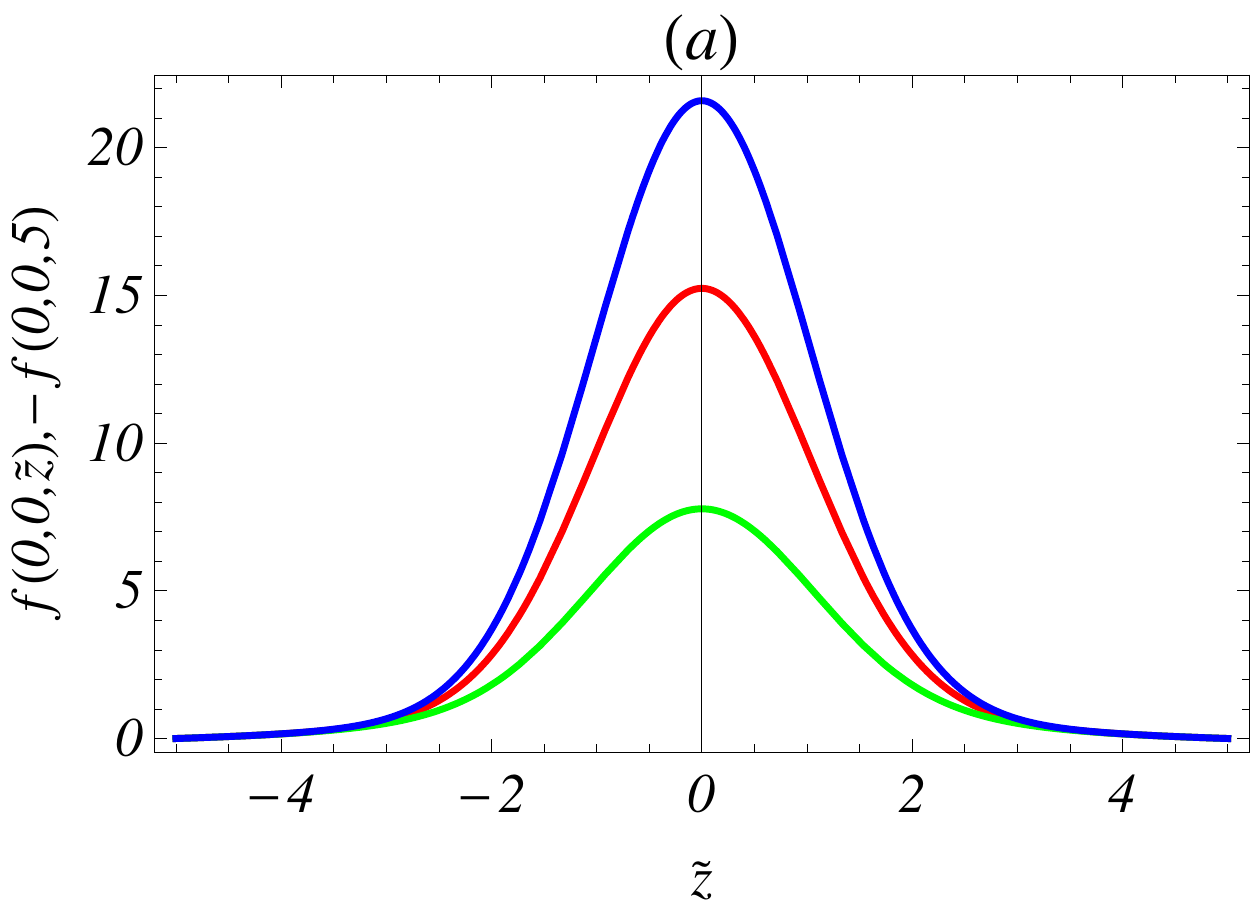}
\hspace{0.2in}
\includegraphics[width=7cm,height=7cm,keepaspectratio]{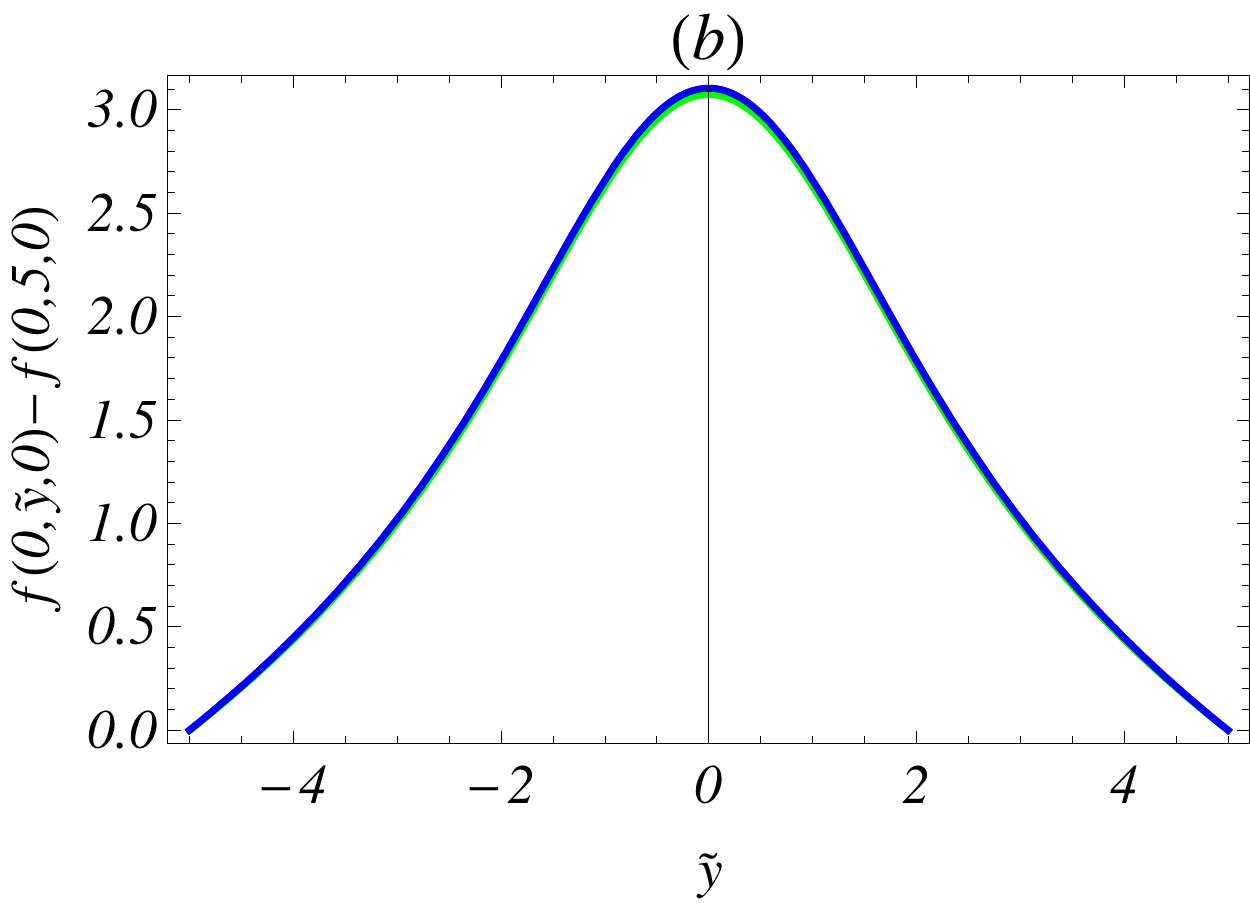}
\caption{Behavior of $f(\tilde{x},\tilde{y},\tilde{z})$  in Eq.~(\ref{frz2}) ) for a cylindrical  bunch ($\eta=1$): (a)  along the $\tilde{z}$ axis  and (b) along the $\tilde{y}$ axis . Blue: $\alpha=10^{-9}$, Red: $\alpha=5.7 \times 10^{-7}$, Green: $\alpha=10^{-3}$.}
\label{faxes} 
\end{figure}

\begin{figure}[h]
\includegraphics[width=7cm,height=7cm,keepaspectratio]{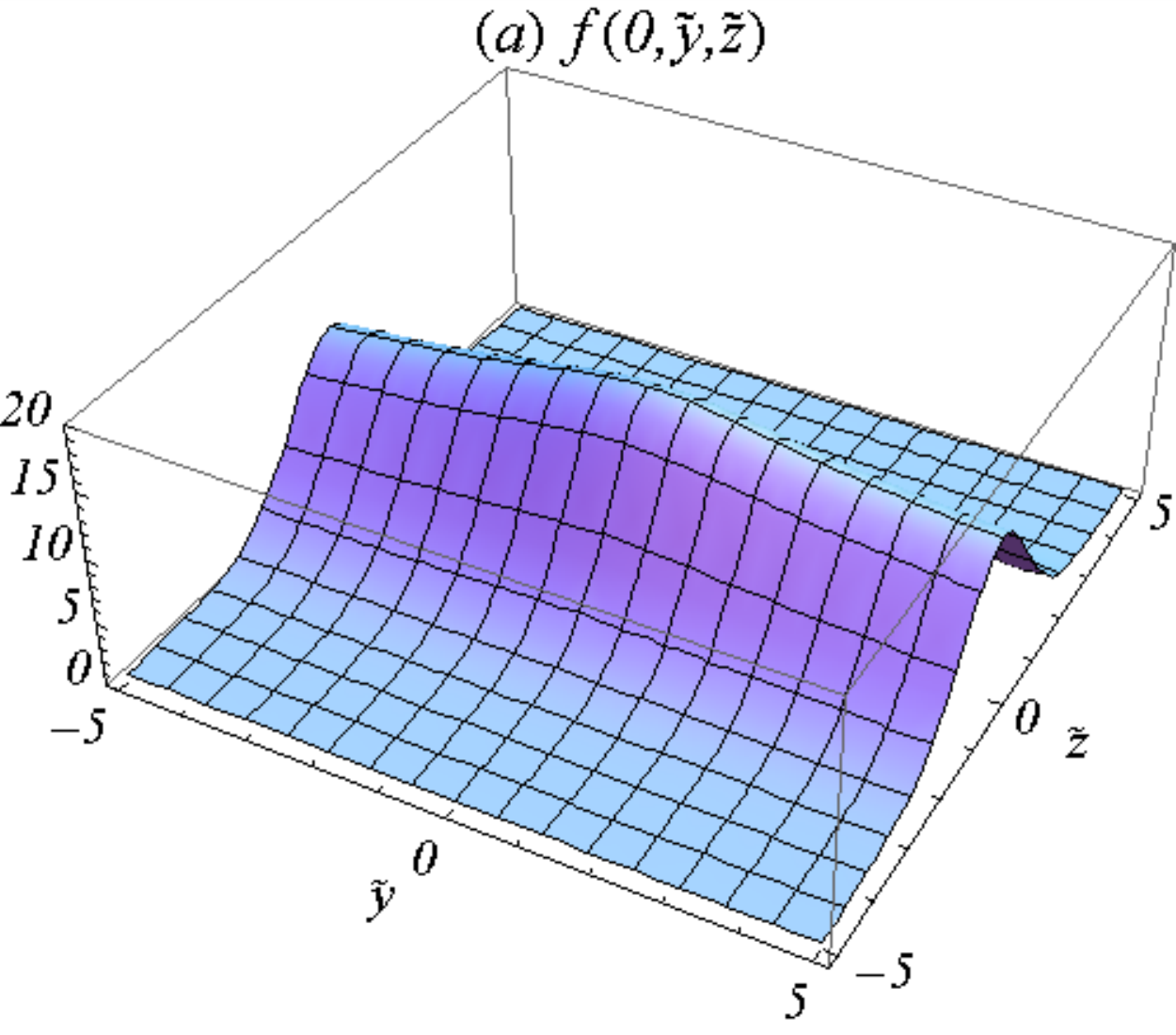}
\includegraphics[width=7cm,height=7cm,keepaspectratio]{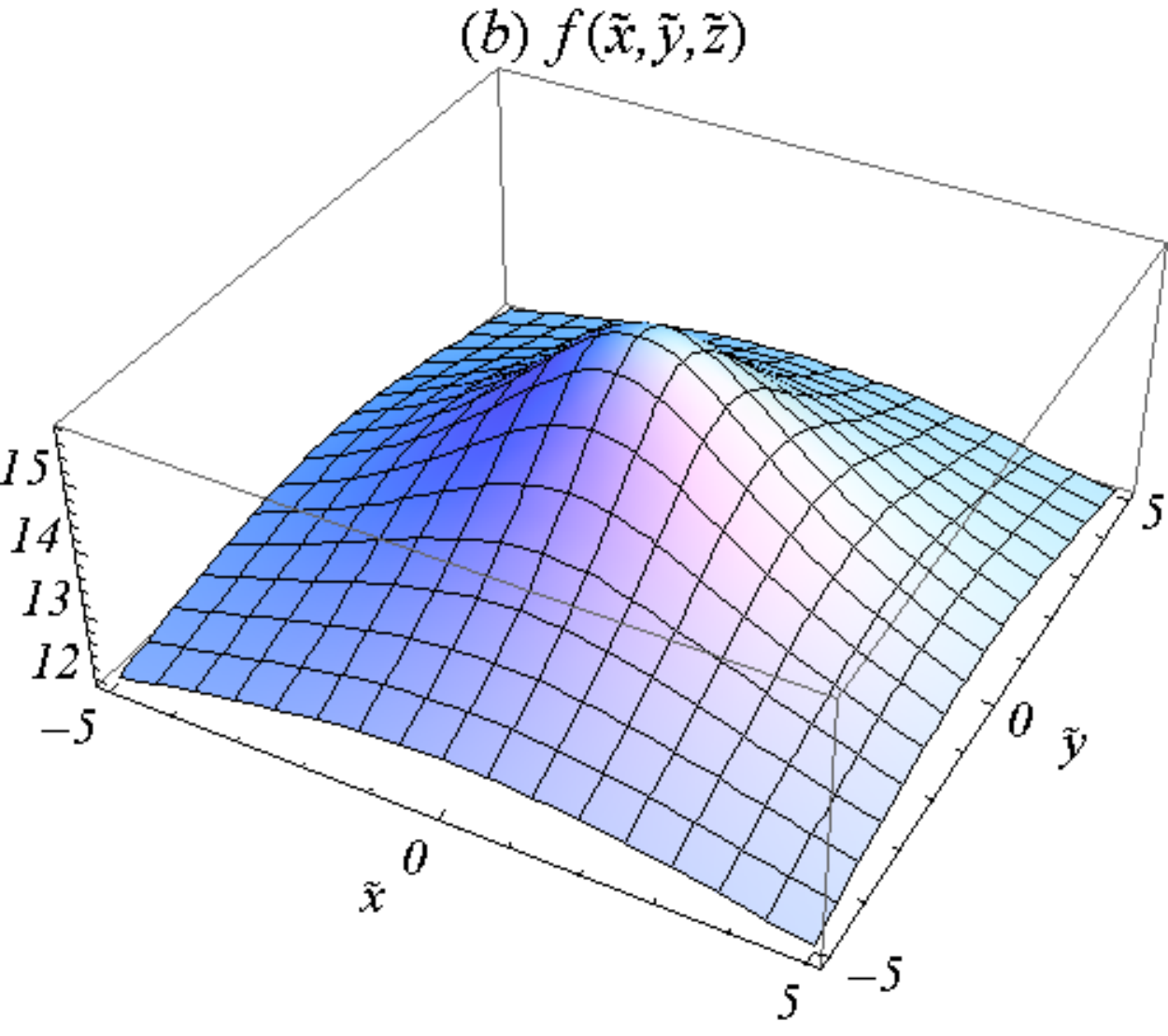}
\caption{Behavior of $f(\tilde{x},\tilde{y},\tilde{z})$  in Eq.~(\ref{frz2})  for a cylindrical  bunch over (a)  the $\tilde{x}=0$ plane and (b)  the $\tilde{z}=0$ plane . Here $\eta=1$ and $\alpha=5.7\times 10^{-7}$  as in Eq.\, (\ref{case1}). }
\label{feta1} 
\end{figure}

\begin{figure}[h]
\includegraphics[width=7cm,height=7cm,keepaspectratio]{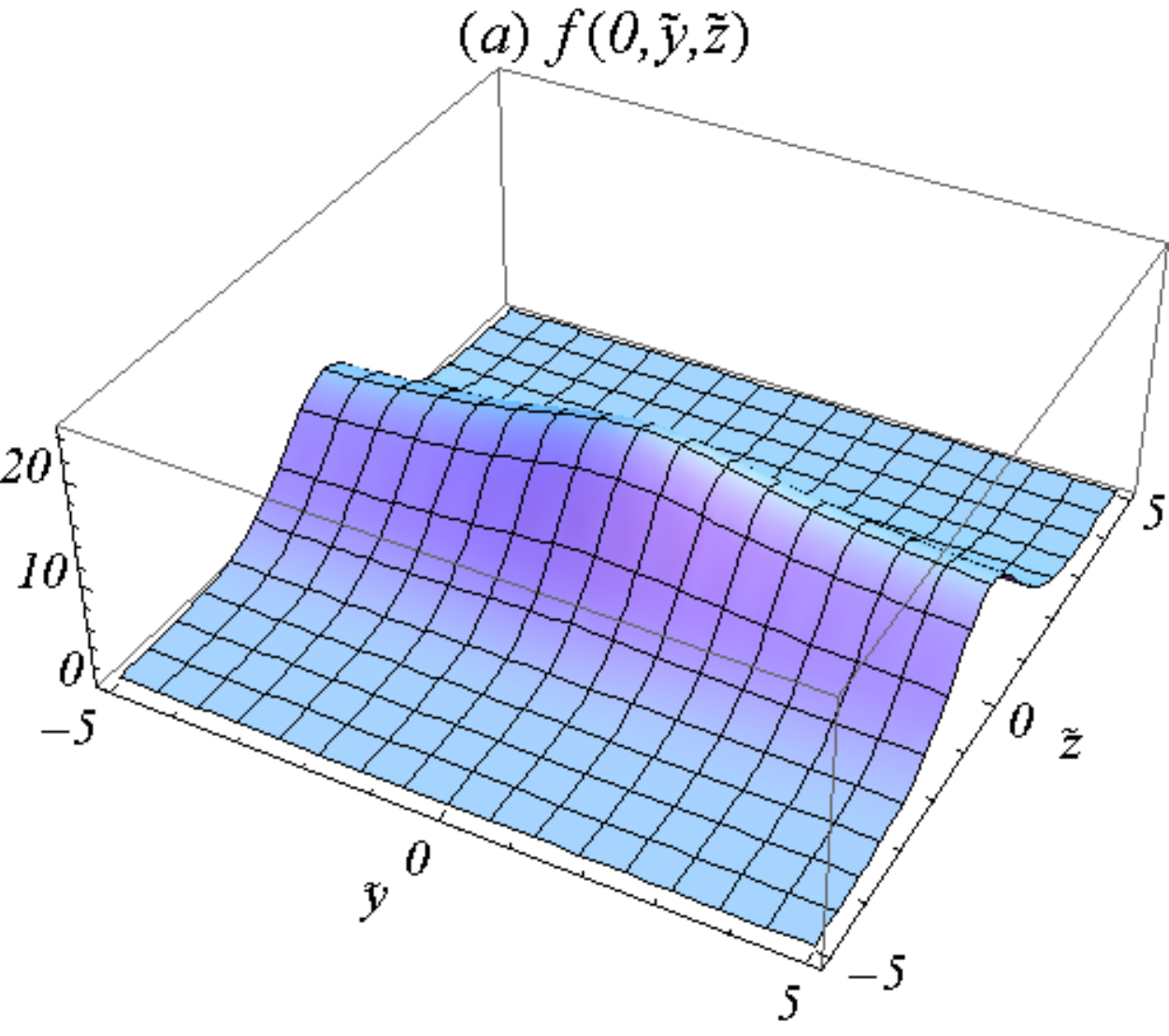}
\includegraphics[width=7cm,height=7cm,keepaspectratio]{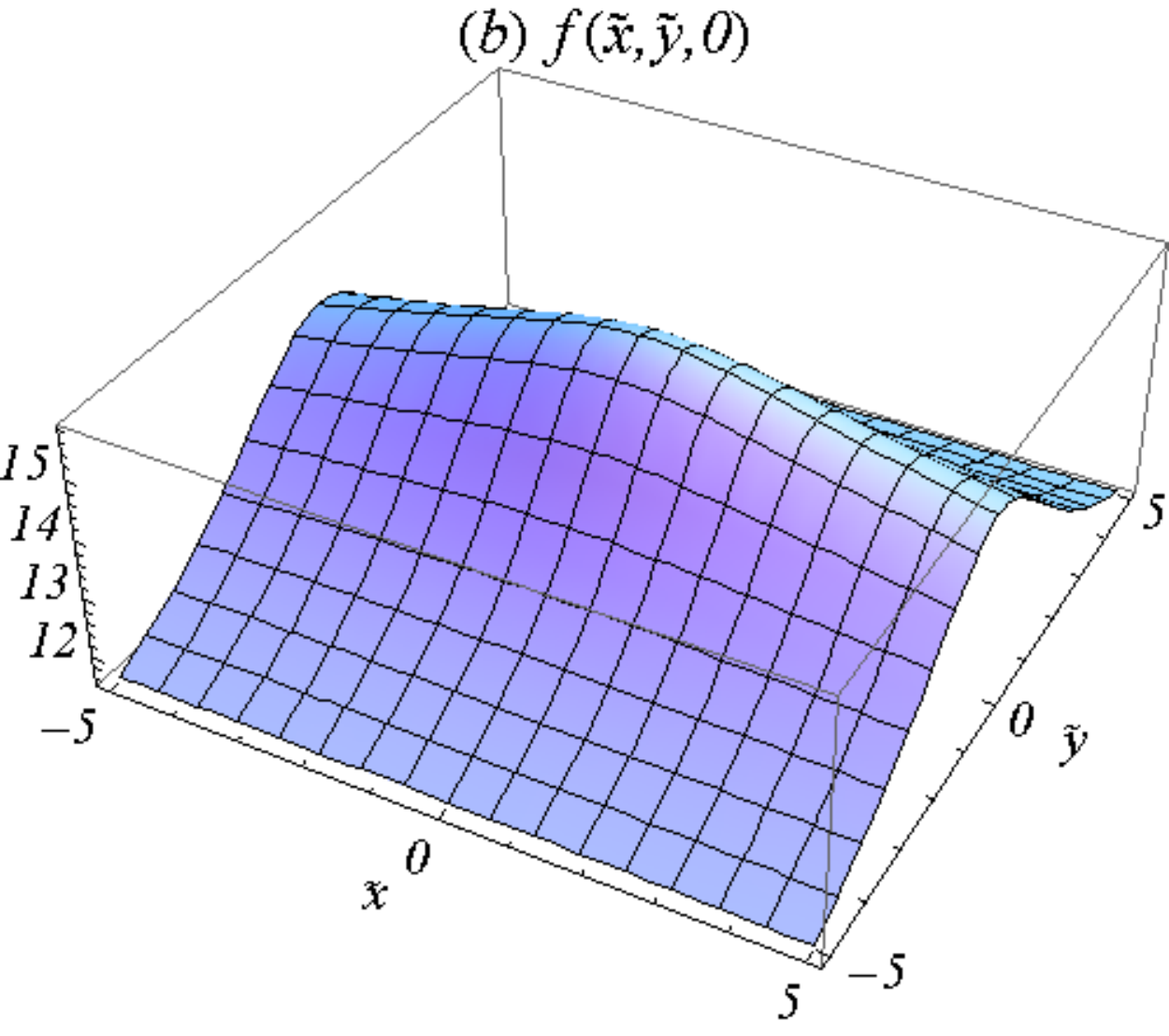}
\caption{Behavior of $f(\tilde{x},\tilde{y},\tilde{z})$  in Eq.~(\ref{frz2}) for a flat  bunch over (a) the $\tilde{x}=0$ plane  and (b)  the $\tilde{z}=0$ plane. Here $\eta=0.02$ and $\alpha=1.7\times 10^{-6}$ as in Eq.\, (\ref{case2}).}
\label{feta02} 
\end{figure}

The expression of the potential energy of an electron at coordinate $(x,y,z)$ within the bunch is given by Eq.~(\ref{efrz})
\beq
\mathcal{E}_{\phi}(x,y,z)\equiv e{\it \Phi}(x,y,z)=\mathcal{E}_{\phi 0} f(\tilde{x},\tilde{y},\tilde{z})   \hspace{0.2in}\mbox{with}\hspace{0.2in} \mathcal{E}_{\phi 0} = mc^2 \frac{I_p}{ I_A},
\label{efrz2}
\eeq
for  $I_p=Nec/(\sqrt{2\pi} \sigma_z)$ being the peak current,   $I_A=e/(r_e c)$=17~kA the Alfven current, and
$f(\tilde{x},\tilde{y}, \tilde{z})$ given by Eq.~(\ref{frz1}) with  $(\tilde{x},\tilde{y},\tilde{z})=(x/\sigma_x, y/\sigma_y, z/\sigma_z)$:
\beq
f(\tilde{x},\tilde{y}, \tilde{z}) = \int_0^\infty \frac{d\tau}{\sqrt{(\tau+\eta) (\tau+1)(\alpha \tau +1)}} 
\exp \left[- \frac{\tilde{x}^2}{2(\tau/\eta +1)}- \frac{\tilde{y}^2}{2(\tau+1)}- \frac{\tilde{z}^2}{2(\alpha \tau +1)} \right].
\label{frz2}
\eeq
Here  $\eta=(\sigma_x/\sigma_y)^2$ and  $\alpha=(\sigma_y/\gamma \sigma_z)^2$.
For a  cylindrical  beam, $\sigma_x=\sigma_y=\sigma_r$, $\eta=1$, and $\tilde{r}^2=\tilde{x}^2+\tilde{y}^2$.  Then $\mathcal{E}_{\phi}(x,y,z)$ becomes $\mathcal{E}_{\phi c}(r,z)$
\beq
\mathcal{E}_{\phi c}(r,z)=\mathcal{E}_{\phi 0} f_c(\tilde{r}, \tilde{z})
\eeq
for
\beq
f_c (\tilde{r},\tilde{z})=\int_{0}^{\infty} \frac{d\tau}{(1+\tau)\sqrt{1+\alpha \tau}}
\exp \left( -\frac{\tilde{r}^2}{2(1+\tau)} -\frac{\tilde{z}^2}{2(1+\alpha \tau)}\right)
\vspace{0.1in}
\label{frz}
\eeq
with  $\tilde{r}=r/\sigma_r$. The potential energy $\mathcal{E}_\phi (r,z)$  reaches its maximum value $\mathcal{E}_{\phi m}$ at $\tilde{r}=\tilde{z}=0$, i.e.,
\beq
\mathcal{E}_{\phi m}=\mathcal{E}_{\phi c}(0,0).
\eeq
The behaviors of the potential energy  in Eq.\,(\ref{efrz2}) for a cylindrical Gaussian bunch along bunch coordinate axes  are  illustrated in Fig.~\ref{faxes}. This plot shows that when $\alpha$ varies, which could be the result  of acceleration \cite{accl} or longitudinal or transverse focusing, the  dependence of  potential energy on $z$  varies while the slice potential energy is insensitive to the variation of $\alpha$. The behaviors of $f(\tilde{x},\tilde{y}, \tilde{z})$  over the $\tilde{y}$-$\tilde{z}$ and $\tilde{x}$-$\tilde{y}$ plane are displayed in Fig.~\ref{feta1}  for  a cylindrical  bunch and in Fig.~\ref{feta02} for a flat bunch.

To quantitatively compare the slice potential energy spread with the usual slice  kinetic energy spread, and in particular, to compute the slice spread of the {\it total} energy, here we use the following parameters for the bunch \cite{lcls, heater}
\beq
 \mathcal{E}_0=135~\mbox{MeV},  \sigma_z=750 \, \mu\mbox{m}, \gamma_0 \epsilon_x=\gamma_0 \epsilon_y=1  \, \mu\mbox{m}.
\label{lcls}
\eeq
with two examples of beta functions at the entrance of a bending system: (i)  cylindrical  bunch with $ \beta_x =  \beta_y = 6  \,\mbox{m} $  and (ii)  flat bunch with $\beta_x = 0.36 \, \mbox{m}, \beta_y = 18 \, \mbox{m}$.  The potential energy spread for the central slice of the bunch can then be estimated using the rigid-bunch formula Eq.\,(\ref{efrz2}) with the following parameters corresponding to the above two cases:
\bea
&& \mbox{case i}: \hspace{0.2in}   \eta=1,\hspace{0.2in}  \alpha =5.7 \times 10^{-7},
\label{case1} \\
&& \mbox{case ii}:   \hspace{0.2in} \eta=0.02,   \hspace{0.2in}\alpha =1.7 \times 10^{-6}.
\label{case2} 
\eea
We will further choose $I_p=120$ A  and thus $\mathcal{E}_0=3.6$\,keV,  and examine situations when  the slice kinetic energy spread takes the typical value $\sigma_{\mathcal{E}_k}=3 \,\mbox{keV}$ \cite{fel08} and a much smaller value $\sigma_{\mathcal{E}_k}$= 1~keV (note that 
for typical LCLS operation the bunch charge is 250 pC, corresponding to $I_p=40$ A with $\sigma_z=750 \, \mu$m).

The calculation of  the spread of $\mathcal{E}_\phi$ requires the knowledge of  the probability distribution of particles over $\mathcal{E}_\phi$, i.e., $P_{\mathcal{E}_\phi}(\mathcal{E}_\phi)$. Consider  $z=0$ slice of the cylindrical  bunch, with the transverse probability distribution 
\beq
P(x,y)dxdy=\frac{1}{2\pi \sigma_x \sigma_y} \exp \left(-\frac{x^2}{2\sigma_x^2}-\frac{y^2}{2\sigma_y^2}\right)\,dxdy
=\frac{1}{2\pi \sigma_r^2} \exp \left(-\frac{r^2}{2\sigma_r^2}\right) rdrd\phi.
\eeq
For $w=\tilde{r}^2$, the probability for particles to lie between $w$ and $w+dw$ is 
\beq
P_w (w)=\frac{1}{2}e^{-w/2},
\hspace{0.2in}\mbox{with}\hspace{0.2in}
\int_{0}^{\infty}P_w(w)dw=1.
\eeq
Then with the one to one correspondence between $w$ and $\mathcal{E}_{\phi}$ following Eq.~(\ref{phic})
\beq
\mathcal{E}_{\phi}(r,0)=\mathcal{E}_{\phi 0}\,U(w),
\label{efw}
\eeq
with $U$ being the normalized potential energy 
\beq
U(w) =\frac{\mathcal{E}_{\phi}}{\mathcal{E}_{\phi 0}}= \int_{0}^{\infty}  \frac{d\tau}{(1+\tau) \sqrt{1+\alpha \tau}} \exp \left( -\frac{w}{2(1+\tau)}\right),
\vspace{0.15in}
\label{uw}
\eeq
one gets the probability for the value of potential energy of a particle to reside between $\mathcal{E}_\phi$ and $\mathcal{E}_\phi + d\mathcal{E}_\phi$  
\beq
P_{\mathcal{E}_\phi}(\mathcal{E}_\phi)\,d\mathcal{E}_\phi=P_U (U) dU=P_w(w)dw
\eeq
or
\beq
P_{U}(w)\equiv P_{\mathcal{E}_\phi}(\mathcal{E}_\phi) \mathcal{E}_{\phi0}=\frac{P_w(w)}{|dU (w) /dw|}.
\label{pefw}
\eeq

\begin{figure}[h]
\includegraphics[width=7cm,height=7cm,keepaspectratio]{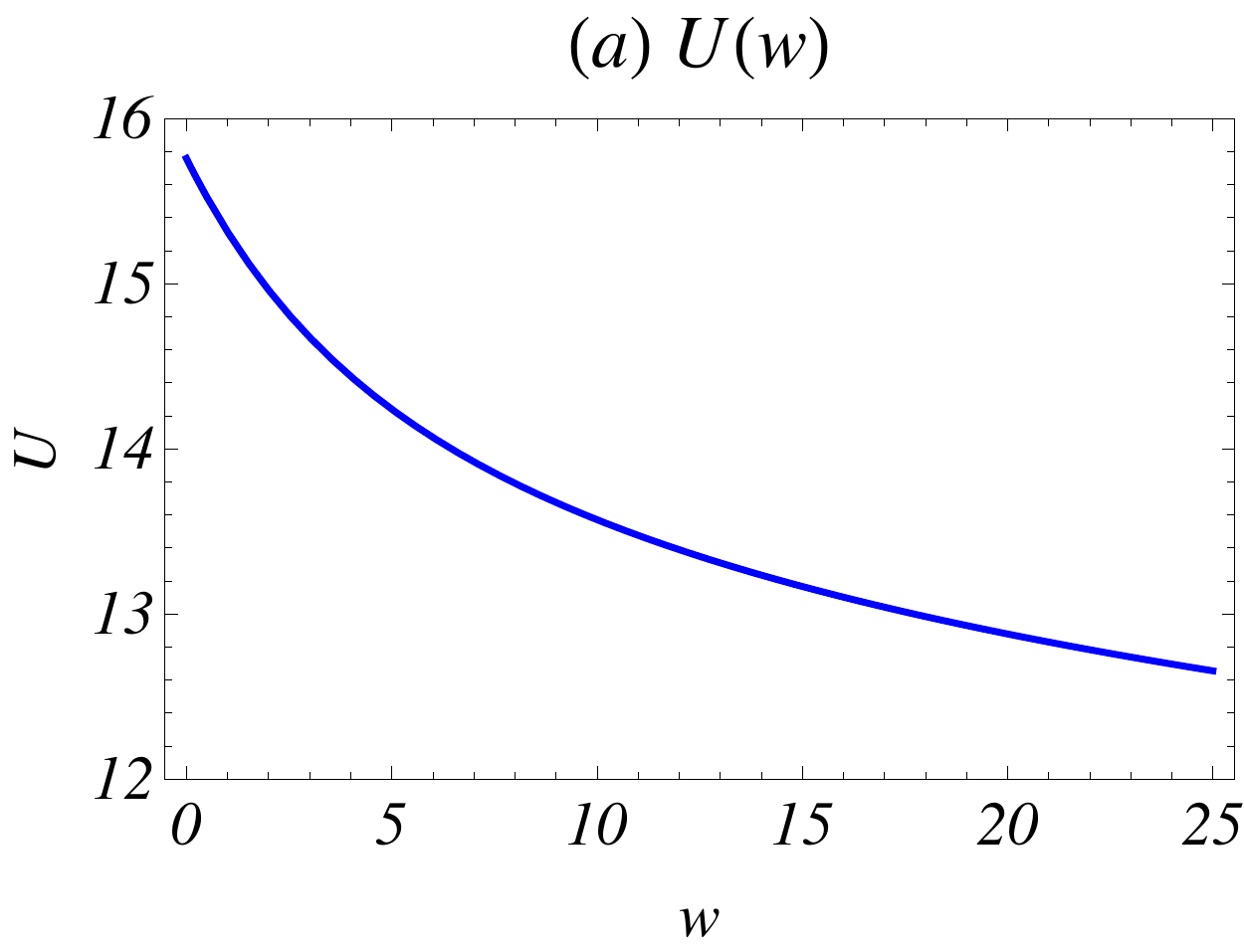}
\hspace{0.2in}
\includegraphics[width=7cm,height=7cm,keepaspectratio]{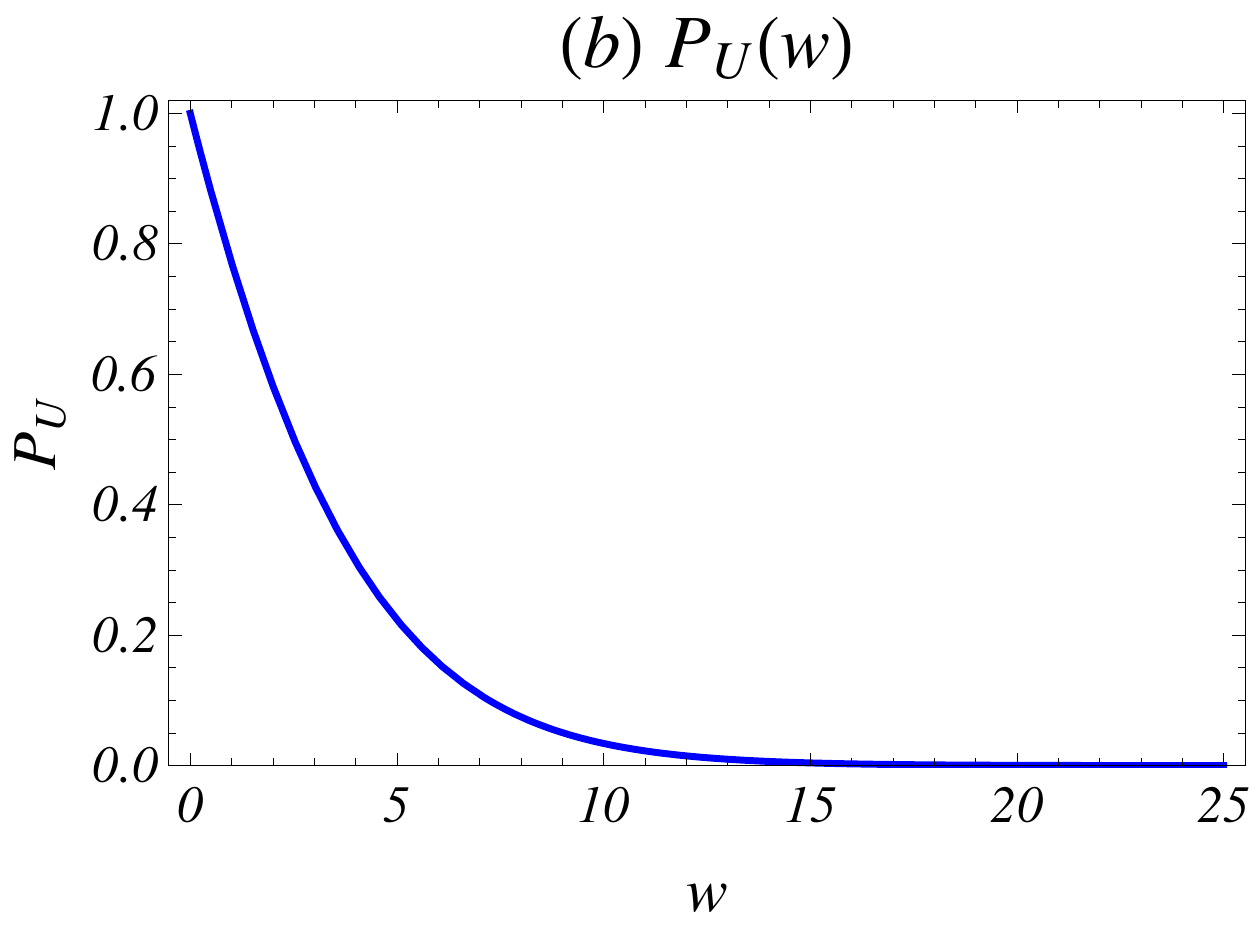}
\caption{Behavior of $U(w)$  in Eq.~(\ref{uw}) and $P_U (w)$ in  Eq.~(\ref{pefw})  for a cylindrical bunch 
 ($\eta=1$, $\alpha=5.7 \times 10^{-7}$):  (a) $U$ vs. $w$ and (b) $P_U$ vs. $w$.}
\label{puw} 
\end{figure}

\begin{figure}[ht]
\includegraphics[scale=0.7]{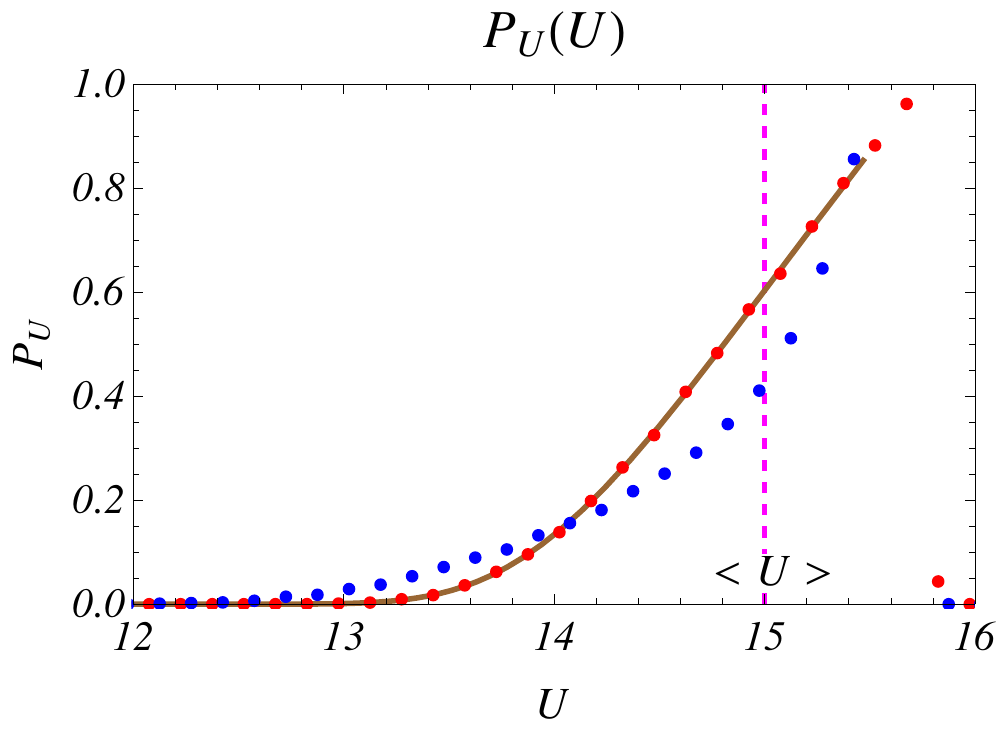}
\vspace{-0.3in}
\caption{$P_ U (U)$  for the central slice of a  Gaussian bunch.  For the cylindrical  bunch  with $\eta=1$ and $\alpha=5.7\times 10^{-7}$, results are shown by the solid brown line (semi-analytical method) and the red dots (Monte Carlo method). For the flat bunch with $\eta=0.02$ and  $\alpha=1.7\times 10^{-6}$, results are shown by  the blue dots  (Monte Carlo method). }
\label{PU} 
\end{figure}

We now look at the probability distribution of potential energy of particles for a cylindrical  Gaussian  bunch. With  $\eta=1$ and $\alpha=5.7 \times 10^{-7}$ as  in Eq.~(\ref{case1}),  we plot   $U (w)$   and  $P_U(w)$ in Fig.~\ref{puw}, which shows that $U$ reaches its maximum value $U_m=U(0)=15.8 $ at $w=0$.
The semi-analytical results of $P_{U}$ on $U$  is shown as the solid brown curve in  Fig.\,\ref{PU}, as obtained  from the parametric dependence of $ (U (w), P_U (w))$ on $w$. The  domain of the function $P_U (U)$  is $(0, U_m)$.
In our study the function $P_U(U)$  is obtained by interpolating an array  $(U (w_i), P_U (w_i))$, with $w_i$ an equally spaced array  in range $ [0,\, 25]$ corresponding to $\tilde{r}$ in the range  $[0,5]$.
Numerically it can  be verified that $P_U(U)$ thus obtained satisfies  
\beq
\int_{0}^{U_ m}P_U (U) dU =1.
\eeq
 Knowing the probability distribution $ P_{U}(U)$, one can further find the average and rms width of the particle distribution over $U$:
\beq
\langle U \rangle=\int_{0}^{U_m} P_U (U) U dU,
\hspace{0.2in}\mbox{and}\hspace{0.2in}
\sigma_{U}=\sqrt{ \langle (\Delta U)^2 \rangle},
\eeq
for
\beq
\langle (\Delta U)^2 \rangle =\int_{0}^{U_m} P_{U}(U) (U- \langle U \rangle)^2  dU.
\eeq
Here $\langle U \rangle=15.0$, and $\sigma_{U}$= 0.51.   Note that the above results have weak dependence on $\alpha$  since $U(w)$ is insensitive to $\alpha$ as indicated  by Fig.\,\ref{faxes}(b).

Finally, we calculate the spread of the total slice energy spread for the $z=0$ slice because  the observed dynamical effect on a bunch  moving  through a bending system is always related to the distribution of the  initial {\it total} energy offset of the particles. The total energy offset from the design energy $\mathcal{E}_0$
for a particle is
\beq
\Delta \mathcal{E}= \Delta \mathcal{E}_k + \mathcal{E}_\phi             
\hspace{0.2in}\mbox{for}\hspace{0.2in}
\Delta \mathcal{E}_k= \mathcal{E}_k - \mathcal{E}_0,
\eeq
with $\mathcal{E}_{\phi}=\mathcal{E}_{\phi 0}\,U$, and  $\mathcal{E}_k$ being the  initial kinetic energy.
Assuming the  probability of the kinetic energy distribution is Gaussian
\beq
P_{\mathcal{E}_{k}}(\Delta \mathcal{E}_k)=\frac{1}{\sqrt{2\pi}\sigma_{\mathcal{E}_{k}}}\exp \left(-\frac{(\Delta \mathcal{E}_k)^2}{2(\sigma_{\mathcal{E}_{k}})^2}\right),
\label{pek}
\eeq
 then the probability of  particle distribution $P_\mathcal{E} (\Delta \mathcal{E})$ over $\Delta \mathcal{E}$ is related to $P_{\mathcal{E}k}(\Delta \mathcal{E}_k)$
and $P_U(U)$ by \cite{rdm}
\bea
P_\mathcal{E} (\Delta \mathcal{E}) &=&\int_{0}^{\mathcal{E}_{\phi m}} P_{\mathcal{E}_{k}}(\Delta \mathcal{E}-\mathcal{E}_{\phi})P_{\mathcal{E}_\phi}(\mathcal{E}_{\phi})\,d\mathcal{E}_{\phi} \nonumber \\
&= & \int_{0}^{U_m} P_{\mathcal{E}_{k}}(\Delta \mathcal{E}-\mathcal{E}_{\phi 0} U) P_U(U)\,dU.
\label{pe}
\eea
The average $\langle \mathcal{E} \rangle$  and root mean square (rms) $\sigma_{\mathcal{E}}$ of $P_\mathcal{E}(\Delta \mathcal{E})$ distribution can be further calculated. An estimation of  $\sigma_{\mathcal{E}}$ is given by
\beq
\bar{\sigma}_{\mathcal{E}} \simeq \sqrt{(\sigma_{\mathcal{E}_k})^2+(\mathcal{E}_{\phi 0}\, \sigma_U)^2}.
\label{sgm}
\eeq
This relation shows that the  rms of the joint energy can be appreciably larger than that of the kinetic slice energy spread  when 
\beq
\xi \equiv \frac{\mathcal{E}_{\phi 0}\, \sigma_U}{ \sigma_{\mathcal{E}_k}} \geq 1,
\label{cri}
\eeq
which often implies  high peak current $I_p$ and small kinetic energy spread $ \sigma_{\mathcal{E}_k}$.

\begin{figure}[h]
\includegraphics[width=7cm,height=7cm,keepaspectratio]{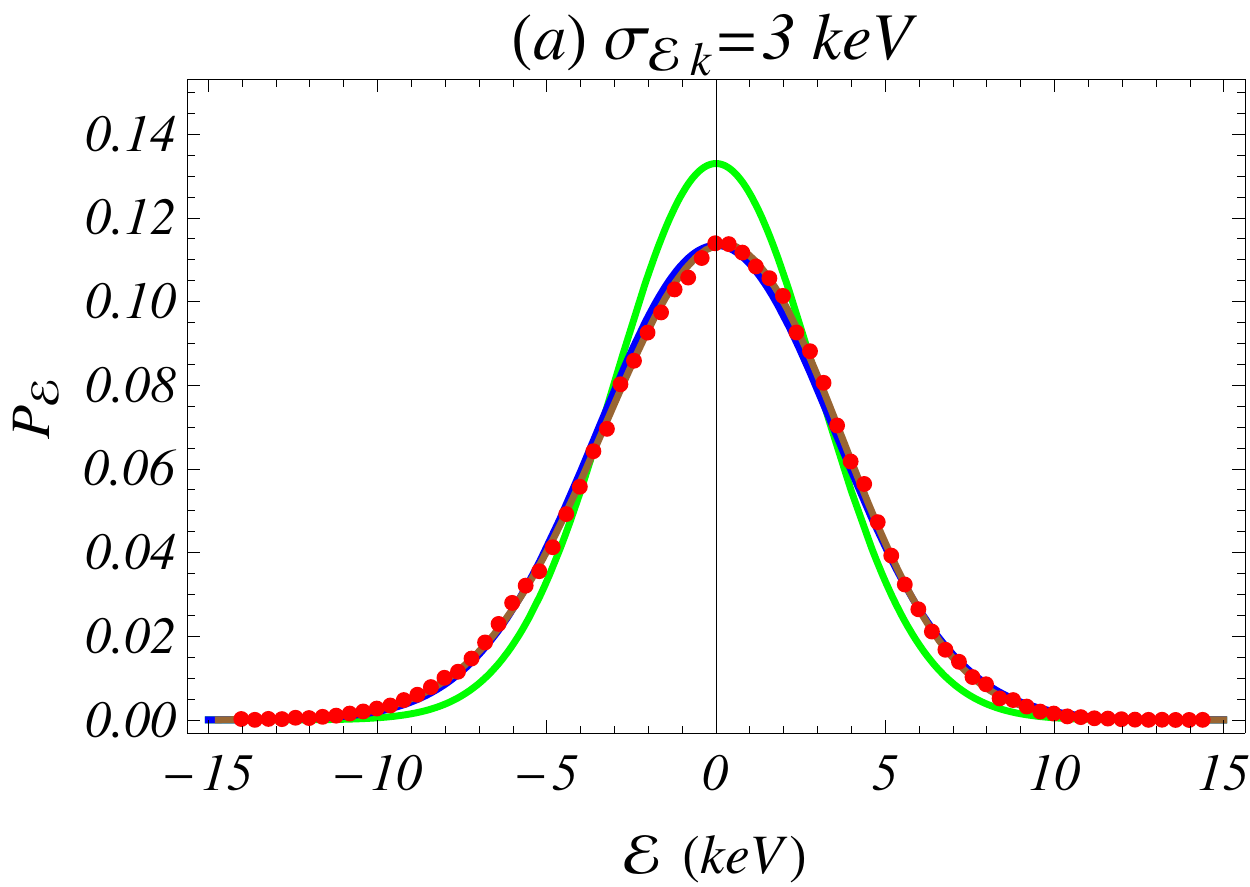}
\hspace{0.3in}
\includegraphics[width=7cm,height=7cm,keepaspectratio]{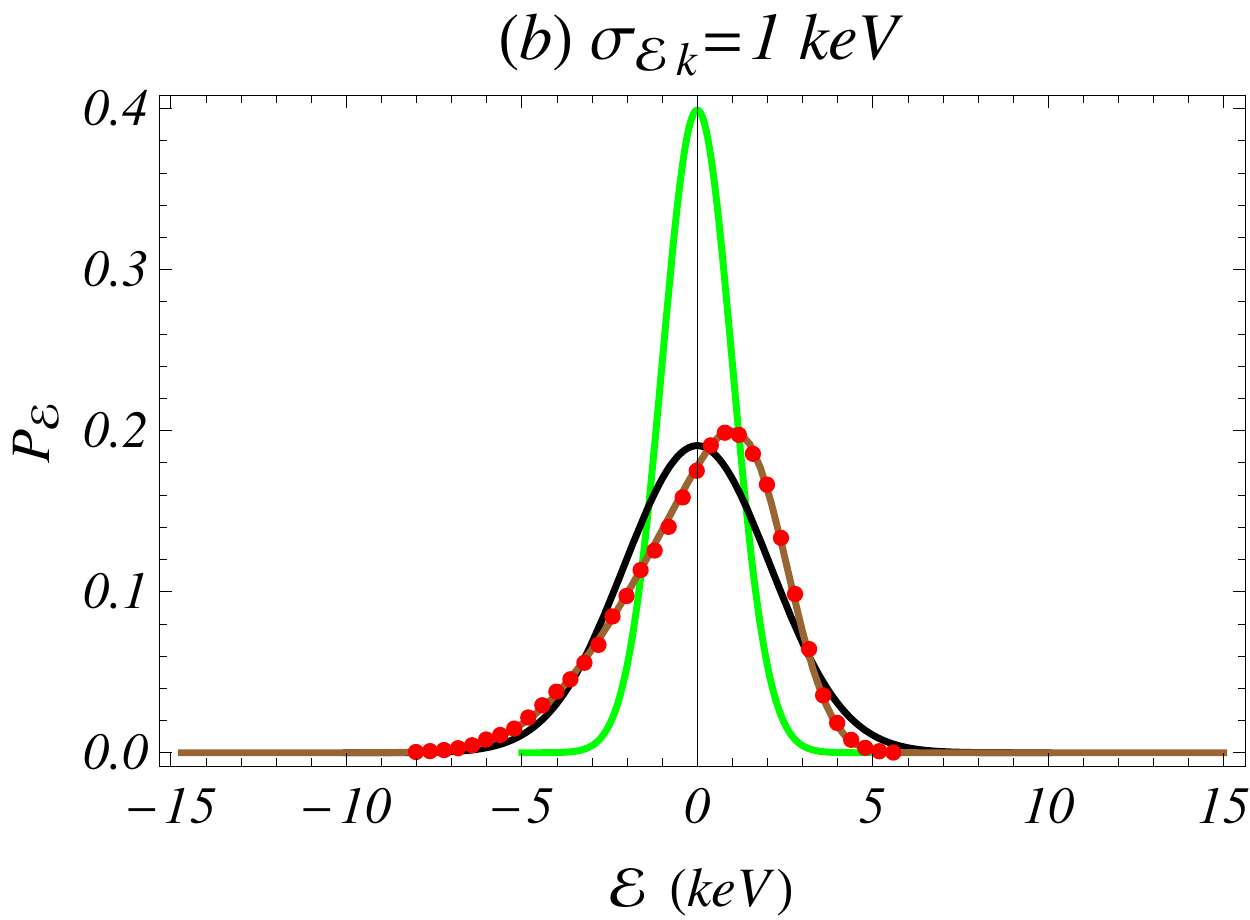}
\caption{ $P_\mathcal{E} (\Delta \mathcal{E})$   for the central slice of a cylindrical  Gaussian bunch with $\mathcal{E}_{\phi 0} = 3.6$\, keV, $\eta=1$, $\alpha=5.7\times 10^{-7}$ for (a) $\sigma_{\mathcal{E}}=3$\, keV and (b) $\sigma_{\mathcal{E}}=1$\, keV. Green line: Gaussian  distribution  for $\Delta \mathcal{E}_k$; Brown line: semi-analytical results of  probability distribution for $\Delta \mathcal{E}$; Black line: Gaussian  distribution  for $\Delta \mathcal{E}$ with estimated rms in Eq.\,(\ref{sgm}); Red dots: Monte Carlo results of  probability distribution  for $\Delta \mathcal{E}$. }
\label{pE} 
\end{figure}

For the parameters in Eq.\,(\ref{case1}),  the probability distribution of the joint energy offset $P_\mathcal{E} (\Delta \mathcal{E})$ can be obtained from  Eq.\,(\ref{pe}) 
by applying  $P_{\mathcal{E}_{k}}(\Delta \mathcal{E}_k)$ in Eq.\,(\ref{pek}) and  the semi-analytical result of $P_U(U)$. The final results are 
shown in Fig.\,\ref{pE} for two cases, $\sigma_{\mathcal{E}_k}= 3 $\, keV for  Fig.\,\ref{pE}(a) when $\xi=0.6 $,  and $\sigma_{\mathcal{E}_k}= 1 $\, keV for  Fig.\,\ref{pE}(b) when  $\xi=1.8 $ . As expected the latter case demonstrates clear effect of slice potential energy spread on the widening of the slice total energy spread. 
Here the solid green lines are the Gaussian probability distribution for the slice kinetic energy offset $\Delta \mathcal{E}_k$ as given by
 Eq.~(\ref{pek}), the solid brown  lines are the semi-analytical results of probability distribution for the joint energy offset as given by Eq.~(\ref{pe}), and the solid black lines are the  Gaussian distribution using the estimated rms in Eq.~(\ref{sgm}).  As an  example,  for  $\sigma_{\mathcal{E}_k}= 1 $\, keV in  Fig.\,\ref{pE}(b),  we find the rms of $P_\mathcal{E} (\Delta \mathcal{E})$ is $\sigma_{\mathcal{E}}=2.09$~keV, which agrees with the  approximation by Eq.~(\ref{sgm})  (for $\mathcal{E}_{\phi 0}=3.6$\, keV and $\langle U\rangle=0.51$). The asymmetric feature of $P_\mathcal{E}(\Delta \mathcal{E})$, as apparently shown by the solid brown lines in  Fig.\,\ref{pE}(b), can be explained by the asymmetric feature  of $P_U (U)$ about $\langle U \rangle$ in Fig.\, \ref{PU}, as the result  of the fact that particles are more likely to populate at smaller values of $r$ where $U$ takes larger value ($U$ is always positive).  

\begin{figure}[h]
\includegraphics[scale=0.7]{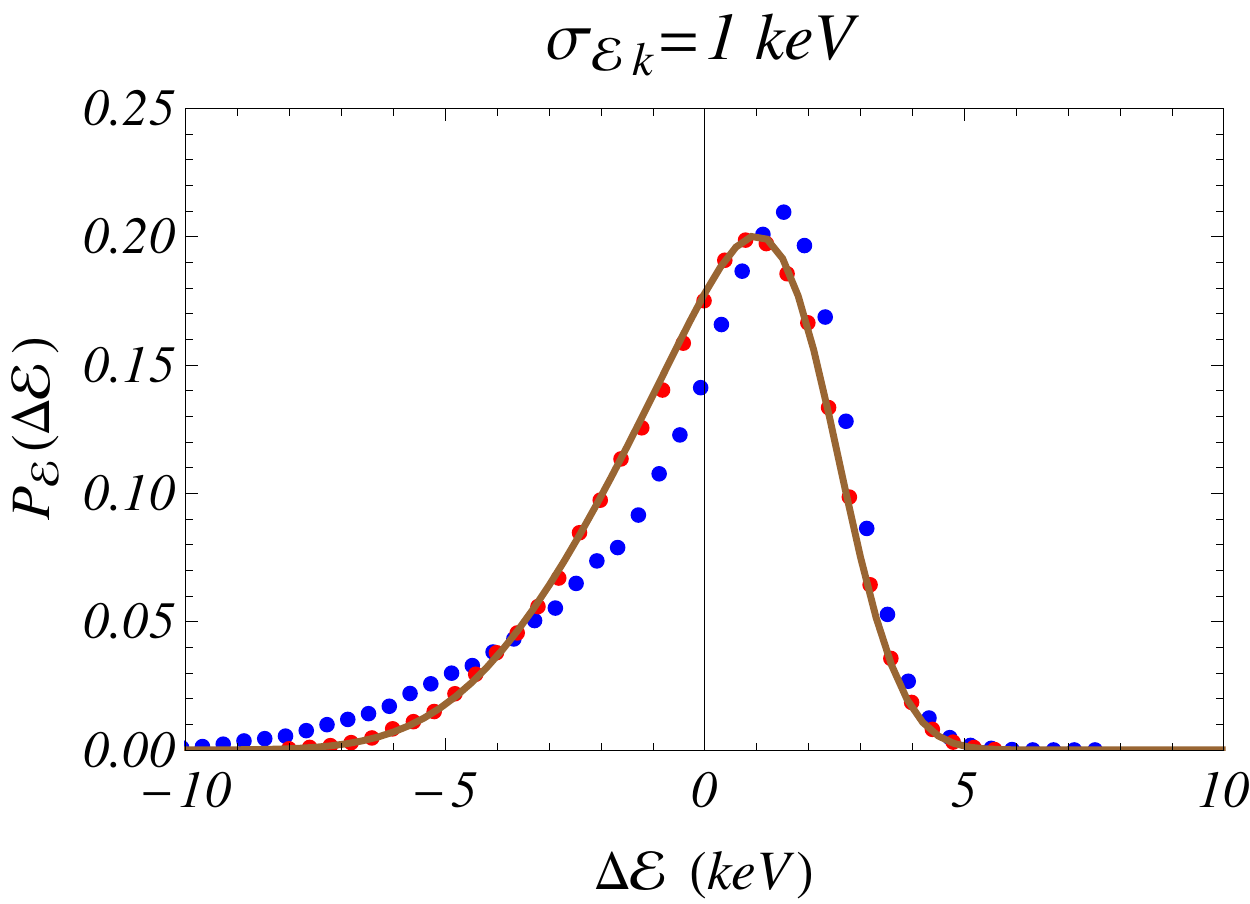}
\caption{ $P_\mathcal{E} (\Delta \mathcal{E})$  for the central slice of a  Gaussian bunch with $\sigma_{\mathcal{E}}=1$\, keV and $\mathcal{E}_{\phi 0} = 3.6$\, keV.  The solid brown line is for semi-analytical result and the red dots are for the  Monte Carlo result ($\eta=1, \alpha=5.7\times 10^{-7}$). The blue dots are for the Monte Carlo result  of the flat bunch case ($\eta=0.02, \alpha=1.7\times 10^{-6})$.
}
\label{data} 
\end{figure}

For a cylindrical Gaussian bunch, the semi-analytical results for the probability distributions of the potential and total energy of the particles, shown in Fig.\,\ref{PU} and Fig.\,\ref{pE}  respectively,  can be verified by the Monte Carlo  approach. In this approach we populate $N$ particles in the 2D configuration space $(\tilde{x},\tilde{y})$ for the $z=0$ slice and in the  kinetic energy offset $ \Delta \mathcal{E}_k$ with random Gaussian distribution. The potential energy  of each particle can be evaluated from Eq.~(\ref{efrz2}) by $\mathcal{E}_{\phi}^i=\mathcal{E}_{\phi }(x^i,y^i,z^i=0)$, and subsequently the total energy offset of the particle is $\Delta \mathcal{E}^i = \Delta \mathcal{E}_{k}^{i} + \mathcal{E}_\phi^{i}$, with superscript $i$ denoting the $i$-th particle. One then finds $P_U (U)$ and $ P_\mathcal{E} (\Delta \mathcal{E})$ from the histogram of $\mathcal{E}_\phi^{i}$ and $\Delta \mathcal{E}^i$ for all particles.  Same approach can also be applied to flat bunch case in Eq.~(\ref{case2}). Here we set $N=100000$. For cylindrical  bunch as in Eq.~(\ref{case1}), the Monte Carlo results are shown by  the red dots in Fig.\,\ref{PU} for $P_{U} (U)$  and   in Fig.\,\ref{pE}  for  $ P_\mathcal{E} (\Delta \mathcal{E})$,  which are all in  good agreement with the semi-analytical results depicted by the solid brown curves in Figs.\, \ref{PU} and \ref{pE}.  For a flat bunch with $\eta=0.02$, the Monte Carlo results are shown as the blue dots in Fig.\,\ref{PU} and Fig.\,\ref{data}.
Here part of Fig.\,\ref{pE}(b) is replotted in Fig.\,\ref{data} for $\eta=1$ and compared with the results for $\eta=0.02$. This comparison shows
that the probability distribution for total energy of particles is more asymmetric for a flat bunch than that for a cylindrical  bunch when Eq.~(\ref{cri}) is satisfied.

\section{\label{sec:dscl}Discussions}

For an electron bunch being transported through a magnetic bending system, we have shown in Sec.\,\ref{sec:eqCM} that the energy spread of the bunch observed in dispersive regions and the  bunch length determined by momentum compaction of the  bending system are actually related to the spread of the {\it total} (or canonical) energy $\mathcal{E}= \mathcal{E}_k +\mathcal{E}_\phi$ of the particles in the  bunch  instead of the spread of the {\it kinetic} energy $ \mathcal{E}_k$ of the particles alone as previously assumed. The spread of $\mathcal{E}_\phi$ is called {\it pseudo energy spread} since the measurements may give the appearance of a larger 
kinetic energy spread as a  result of the pseudo energy spread.
In  Sec.\,\ref{sec:eFi},  by analyzing the probability distribution $P_\mathcal{E} (\Delta \mathcal{E})$   of the total energy offset from design energy for the central ($z=0$) slice of a rigid cylindrical Gaussian bunch moving relativistically on a straight path,  we find that  the contribution of ${\mathcal{E}_\phi}$ to the total energy spread can be  appreciable when  Eq.\,(\ref{cri}) is satisfied, i.e., when a bunch has  high peak current and low slice kinetic energy spread.
In this section, the implication of pseudo slice energy spread for bunch dynamics in bends, including microbunching instability, will be discussed. Other effects not included in this study that requires further consideration will also be highlighted.

\subsection{Effects Not Included in this Study}

First, the present study assumes free-space boundary condition.  The existence of wave-guide boundary will alter the functional form of $\mathcal{E}_\phi (x,y,z)$  by adding a solution of homogeneous wave equation to the free-space potential. Despite this modification, the sensitive dependence of $\mathcal{E}_\phi$ on $(x,y)$  will  preserve since it originates from local interaction.

Second,  this study emphasizes  on the driving term $G_{\phi 0}$ in Eq.\,(\ref{Gf0}), yet the impact of {\it effective} CSR forces, which   tend to generate correlated (with $z$) phase space distortions in bending systems, are not considered.
For 1D steady-state CSR  interaction, one has $d\Phi^{\mbox{\scriptsize col}}/dt=0$ and $dA_r^{\mbox{\scriptsize col}}/dt=0$, and the behavior of  $F_{v}^{\mbox{\scriptsize eff}}$  and  $F_{r}^{\mbox{\scriptsize eff}}$ are analyzed 
\cite{derb2}, with the effect of  $F_{v}^{\mbox{\scriptsize eff}}$ on transverse dynamics dominant over 
that of  $F_{r}^{\mbox{\scriptsize eff}}$  \cite{hbb}.
A study of the effect of 2D dynamics \cite{lir08} on longitudinal effective CSR force $F_{v}^{\mbox{\scriptsize eff}}$  shows that using an actual evolving bunch rather than a rigid line bunch could lead to a delayed response of $F_{v}^{\mbox{\scriptsize eff}}$ to the bunch length variation. In addition,  $F_{v}^{\mbox{\scriptsize eff}}$  could be sensitive to the transverse particle coordinates  for a  short duration around roll-over compression, as a result of sensitivity of longitudinal phase to  transverse particle positions when Derbenev criterion \cite{derb2} is not satisfied.  So the resulting integrated effect of   $F_{v}^{\mbox{\scriptsize eff}}$  can have transverse dependence only after role-over compression. The general effect of $F_{r}^{\mbox{\scriptsize eff}}$, including the contribution from  $dA_r^{\mbox{\scriptsize col}}/dt$ term,  requires further investigation. Presently the measured emittance growth  in chicanes show good agreement with start-to-end simulations using Elegant \cite{bane}. This does not contradict with our newly introduced $\mathcal{E}_\phi$ term, since just as the initial slice kinetic energy spread, the $\mathcal{E}_\phi$ term does not cause emittance growth in an achromatic bending systems.

Third, in general,  $P_\mathcal{E} (\Delta \mathcal{E})$ depends on the 3D density distribution of the bunch and the history of bunch 3D dynamics.  In actual machines the bunch phase space distribution can have complicated structures, as seen in many start-to-end simulations \cite{tesla, ding}. Thus detailed studies of the  dependence of potential energy on the spatial coordinates are needed for each experimental setting, which may show  behavior of the potential energy distribution very different from that for a perfect 3D Gaussian bunch as discussed in this paper.  Note  that our focus here is on the pseudo slice energy spread for the central slice of the Gaussian bunch,  and its dependence  on the longitudinal position of the slice is not considered in this paper (it  gets smaller for larger $z$ as shown in Figs.\,\ref{feta1}(a)  and \ref{feta02}(a)).   Likewise,
 the maximum value  of potential energy $\mathcal{E}_{\phi m}$ for each slice varies with $z$, rendering a curvature of $\mathcal{E}_{\phi m} $ vs. $z$ similar to the effect of  RF curvature on beam longitudinal phase space distribution. It remains to be examined  whether  the curvature of potential energy is always negligible compared to the RF curvature in experiments of interest.

Finally, one should note that in an ideal numerical simulation the cancellation is realized by taking full account of the transverse and longitudinal Lorentz forces, including their local and long range behaviors, and let the dynamics play out with full self-consistency. In this way the interplay of effects from the longitudinal and transverse forces can be faithfully represented, the cancellation can be manifested, and the effect of the  pseudo (kinetic) energy term  can be revealed.  Such interplay may often misrepresented if only partial effects are included. 

\subsection{Roles of $\mathcal{E}_{\phi}$ in Experimental Measurement}

As we have discussed,  the effects of the pseudo slice energy spread are measurable  in experiments  only for
special cases when the bunch peak current is high and the slice kinetic energy spread is low. For a rigid Gaussian bunch the criterion is given by Eq.\,(\ref{cri}).
In most of the  existing experiments, the parameters are such that this effect may  not  be significant. Nonetheless, as the community is pushing for higher peak current
and higher brightness of electron beams, this effect may show up in certain future experiment. Here we make some general comments of  its possible impact on bunch dynamics in bending systems.

A common scheme for measuring the slice energy spread is to use the transverse deflecting  RF structure (TDS) together with a spectrometer \cite{tds}. According to the Panofsky-Wenzel theorem, TDS can introduce additional local energy spread to the beam.  Further evaluations are required to check if  the potential energy distribution over 
the streaked bunch at the entrance of the spectrometer can cause detectable perturbation to  the measurement  of the  slice kinetic  energy spread.

Recently measurement of local energy spread was successfully carried out using the coherent harmonic generation (CHG) method \cite{chg}. In this scheme, an electron bunch is first modulated by $\Delta E=\Delta \gamma  \sin z$
with a seeded laser in an undulator, and then is sent through a chicane where energy modulation is converted  to density modulation. A second undulator is followed in which microbunched beam generates powerful coherent harmonic radiation, with CHG power $E_{\mbox{\scriptsize CHG}}$  related to the bunching factor $b_n$ by  $E_{\mbox{\scriptsize CHG}} \propto b_n^2$. Here the bunching factor at the exit of the chicane is obtained from particle's coordinates $(x_0,y_0,z_0)$ at the entrance of the chicane via
\beq
z=z_0+R_{56} \left (\frac{\Delta \gamma \sin z}{\gamma} +\frac{\mathcal{E}_k}{\gamma mc^2} \right).
\eeq
 However, with our present results of $\mathcal{E}_\phi$  effect on longitudinal optics, we will  have
\beq
z=z_0+R_{56} \left (\frac{\Delta \gamma \sin z}{\gamma} + \frac{\Delta \mathcal{E}_k +\mathcal{E}_\phi (x_0,y_0,z_0)}{\gamma mc^2}\right),
\label{zz0}
\eeq
and the bunching factor  can be obtained by including additional integration over transverse density distribution. For the recent measurement \cite{chg}, with bunch charge 100\,pC and (FWHM) pulse length 8\,ps, we have $\mathcal{E}_{\phi 0}= 0.35 $\, keV or $\xi \sim 0.18$ . Therefore perturbation on the kinetic energy spread  ($\sigma_k \sim 1 $\,keV) from $\mathcal{E}_{\phi}$ is negligible if a rigid cylindrical  Gaussian bunch is used for a rough estimation. In general, magnetic chicanes are commonly used in the CHG  experiments as well as the HGHG and  EEHG \cite{eehg}  experiments, in which slice  energy spread is a crucial parameter.  
In case the bunch peak current is pushed to the point when Eq.\,(\ref{cri}) is satisfied, special care is required to evaluate the effects of  $\mathcal{E}_\phi$ on the final  bunching factor and efficiency of these schemes for  high harmonic numbers.

At high peak current, the pseudo slice energy spread $\mathcal{E}_\phi$ can also have impact on the longitudinal space charge induced microbunching instability, developed  on a beamline consisting of a straight section followed by a chicane. Note
that even for cases with the kinetic energy spread comparable with the potential energy spread,  the contributions
of $\mathcal{E}_k$ and $\mathcal{E}_\phi$ to the final bunching factor can be quite different, because  $\exp (ik R_{56} \mathcal{E}_\phi(x_0,y_0,z_0)/\mathcal{E}_0)$  in Eq.~(\ref{zz0}) will be averaged over transverse spatial distribution, yet $\mathcal{E}_{k}$ is  averaged over the  energy distribution.

The emphasis of this paper is that the pseudo slice energy spread plays an equal role with the kinetic energy spread in their contribution to the measured values of the energy spread in dispersive regions and to the minimum bunch length at full compression when the bunch is transported through a magnetic bending system. It would be interesting to demonstrate this effect of pseudo slice energy spread with clarity by creating an experimental condition of high peak bunch current and low slice kinetic energy spread. Such condition may require an
emittance exchange from the longitudinal phase space to the transverse phase space, which is a process reversed from the usual practice,
namely,  converting the transverse phase space emittance to the longitudinal one \cite{eex}. It is expected that transporting such a beam with low  longitudinal phase space emittance  through a bending system will  allow the pseudo slice energy spread  to have  pronounced effects in measurements.

It should be noted that many discussions in this paper assume a random Gaussian distribution for the kinetic energy of the particles, while the potential energy of the particles varies depending on the particle coordinates inside the bunch. However, in actual bunch dynamics, the kinetic and potential energies are closely correlated because their sum---the canonical energy---is  changed by the effective longitudinal force which in most cases \cite{roll} only depends on $z$ at ultrarelativistic limit. The correlation between the kinetic and potential energy of particles can be found  by a self-consistent start-to-end simulation with all space charge and CSR forces included.

In past years there have been puzzles that sometimes the measured slice energy spread is bigger than the expected value obtained from numerical modeling of beam generation from the gun and beam transport in the injector  \cite{uBI}. There could be various reasons causing this puzzle.  Depending on the actual parameters of an experiment, the possible contribution  of pseudo slice energy spread  to the measured value of slice kinetic energy spread needs to be carefully evaluated and sorted out.

\section{\label{sec:conclusion}Conclusion}

In this study, for an ultrarelativistic electron bunch being transported through a magnetic bending system, we consider a remnant driving term of particle transverse dynamics after the cancellation between the Talman's force and the non-inertial space charge force in particle transverse dynamics is taken into account. This driving term is
related to the initial potential energy of the particles at the entrance of the bending system, which has sensitive dependence on the transverse coordinates of particles in the bunch,  and it does not cause emittance growth for an achromatic bending system. Without careful analysis of the detailed interplay of the longitudinal and transverse CSR forces in particle transverse dynamics, the effect of this term may appear experimentally in disguise as a part of the  kinetic energy spread. Our estimation for a cylindrical   Gaussian bunch shows that the slice potential energy spread, or pseudo energy spread, can be comparable in magnitude with the slice  kinetic energy spread when the bunch peak current is high and the slice kinetic energy spread is small. This result renders the importance in sorting out the possible contribution of $\mathcal{E}_\phi$ term from the measured results of slice energy spread. The possible role of $\mathcal{E}_\phi$ in various experimental designs involving chicanes are discussed. Finally, we note that the effect of pseudo slice energy spread is not included in the common 1D CSR simulation, and it can only be revealed by an accurate, complete and fully self-consistent CSR interaction model. Further numerical and experimental studies are needed for reaching a more complete and thorough understanding of the effect of  pseudo slice energy spread.

\begin{acknowledgments}

This work is supported by Jefferson Science Associates, LLC under U.S. DOE Contract No. DE-AC05-06OR23177.

\end{acknowledgments}

\appendix

\section{Retarded Potential}

 We now analyze the retarded scalar potential on particles for a rigid 3D  Gaussian bunch moving on a straight path, and show that
the analytical  expression  of the retarded potential is identical to the scalar potential as obtained by  applying  Lorentz transformation to the 4-vector potentials from the bunch comoving frame to the lab frame. 

Consider an ideal case when a Gaussian bunch moves on a straight path,  with all particles  moving only  longitudinally  and at constant velocity $v=\beta c$. The  motion of the  bunch center is described by $s=\beta ct$. The retarded potential on a test particle at $({\bf x},t)$  is 
\beq
\Phi^{\mbox{\scriptsize col}}({\bf x},t)=\int \frac{\rho({\bf x}_r,t_r)}{|{\bf x}-{\bf x}_r|} d^3 {\bf x}_r,
\label{fxt}
\eeq
where $t_r=t-|{\bf x}-{\bf x}_r|/c$ is the retarded time, and
\beq
\rho({\bf x}_r,t_r)= \frac{Ne}{(2\pi)^{3/2} \sigma_x\sigma_y\sigma_z} \exp \left(-\frac{x_r^2}{2\sigma_x^2}-\frac{y_r^2}{2\sigma_y^2}-\frac{(s_r-\beta ct_r)^2}{2\sigma_z^2}\right).
\eeq
Here $(x_r,y_r,s_r)$ are the Cartesian coordinates of a source particle in the lab frame, and $N$ is the number of electrons in the bunch.

Let $z=s-\beta ct$ be the longitudinal position of the test  particle relative to the bunch center. For $\Delta s=s-s_r$, we have for the source particle
\beq
z_r=s_r-\beta c t_r=z-\Delta s +\beta |{\bf x}-{\bf x}_r|.
\vspace{-0.1in}
\eeq
With the identities \cite{math}
\beq
\frac{1}{\sqrt{2\pi}\sigma_z} \exp \left( -\frac{(s_r-\beta ct_r)^2}{2 \sigma_z^2} \right)
=\frac{1}{\pi} \mbox{Re}\int_{0}^{\infty} 
\exp \left[ -\frac{k^2\sigma_z^2}{2}+i[k(z-\Delta s)+k\beta |{\bf x}-{\bf x}_r|] \right]\,dk
\vspace{0.1in}
\eeq
and 
\beq
\frac{ \exp (ik \beta |{\bf x}-{\bf x}_r|)} { |{\bf x}-{\bf x}_r| }=\frac{1}{\pi}
\int_0^{\infty} \exp \left[  i \left( (k\beta u)^2 + \frac{|{\bf x}-{\bf x}_r|^2}{4u^2} -\frac{\pi}{4} \right) \right] 
\frac{du}{u^2},
\vspace{0.1in}
\eeq
and with the  new variable $w=u^2$, $\Phi^{\mbox{\scriptsize col}}({\bf x},t)$ in Eq.~(\ref{fxt}) becomes  ${\it \Phi}(x,y,z)$ in terms of particle coordinates internal to the bunch
\bea 
{\it \Phi}(x,y,z) & = & 
\mbox{ Re} \left\{\frac{i}{\sqrt{2\pi}} \int_0^{\infty}
\frac{dw}{\sqrt{(\sigma_x^2+iw)(\sigma_y^2+iw) (\sigma_z^2+iw/\gamma^2)  }}  \right. \nonumber \\
& &   \left. \times \exp \left[- \frac{x^2}{2(\sigma_x^2+iw)}- \frac{y^2}{2(\sigma_y^2+iw)}- \frac{z^2}{2(\sigma_z^2+iw/\gamma^2)}\right]
\right\}.
\label{c1}
\eea
Here the path of  integration for $i w$  is $C_1$   in Fig.~{\ref{contour}. For the above integrand,  we have in Fig.~{\ref{contour} $\int_{C_1} + \int_{C_R} +\int_{C_2}=0 $, as well as $ \int_{C_R} =0$ for $R\rightarrow \infty$. This allows us to replace $\int_{C_1}$ by $ -\int_{C_2}$, or
\bea
{\it \Phi}(x,y,z) &= & \frac{ N e}{\sqrt{2\pi}\sigma_z} \int_0^{\infty} \frac{d\tau }{\sqrt{(\tau+\eta) (\tau+1) (1+\alpha \tau)} } \nonumber \\
&& \times \exp \left( -\frac{\tilde{x}^2}{2(1+\tau/\eta)}-\frac{\tilde{y}^2}{2(1+\tau)}-\frac{\tilde{z}^2}{2(1+\alpha \tau)} \right)
\label{phic1}
\eea
for $(\tilde{x},\tilde{y},\tilde{z})=(x/\sigma_x, y/\sigma_y, z/\sigma_z)$, along with $\eta=(\sigma_x/\sigma_y)^2$ and  $\alpha=(\sigma_y/\gamma \sigma_z)^2$.

Next we show that Eq.~(\ref{phic1}) can be obtained by applying Lorentz transformation on the scalar potential from the bunch comoving frame to the lab frame.  In the comoving frame, let $(x',y',s')$ be the particle coordinates, $\Phi'$  be the scalar potential, and let  the rms bunch length be $\sigma'_x, \sigma'_y$ and $\sigma'_z$.  One has \cite{kheifets}
\begin{multline}
\Phi'({\bf x'})  =  \int d{\bf y'} \frac{\rho({\bf y'})}{|{\bf x'}-{\bf y'}|}   
=\frac{Ne}{\sqrt{\pi}} \int_0^\infty \frac{d u}{\sqrt{(u+2 \sigma_x^2)(u+2 \sigma_y^2)(u+2 \sigma_z^2)}} \\
 \times 
\exp \left(- \frac{x'^2}{u+2\sigma_x^{'2}}- \frac{y'^2}{u+2\sigma_y^{'2}}- \frac{s'^2}{u+2\sigma_z^{'2}} \right).
\label{fip}
\end{multline}
Combining $\Phi'({\bf x'})$ in  Eq.~(\ref{fip}) with the Lorentz transformation, i.e., $\Phi'=\gamma \Phi$, $s'=\gamma (s-\beta ct)\equiv \gamma z, x'=x, y'=y$, together with $\sigma'_z=\gamma \sigma_z, \sigma'_x=\sigma_x, \sigma'_y=\sigma_y$ and  $u=2\sigma_y^2 \tau$, 
one  gets the scalar potential in the lab frame 
\begin{multline}
\Phi(x,y,s,t)  = {\it \Phi}(x,y,z)= \int d{\bf x}_1 \frac{\rho({\bf x}_1)}{|{\bf x}-{\bf x}_1|}   
=\frac{ Ne}{\sqrt{2\pi} \sigma_z} \int_0^\infty \frac{d\tau}{\sqrt{(\tau+\eta) (\tau+1)(\alpha \tau +1)}} \\
 \times 
\exp \left(- \frac{\tilde{x}^2}{2(\tau/\eta +1)}- \frac{\tilde{y}^2}{2(\tau+1)}- \frac{\tilde{z}^2}{2(\alpha \tau +1)} \right)
\end{multline}
which is identical to the retarded potential in Eq.~(\ref{phic1}) as expected.

The potential energy of an electron at coordinate $(x,y,z)$ within the bunch is 
\beq
\mathcal{E}_{\phi}(x,y,z)\equiv e{\it \Phi}(x,y,z)=\mathcal{E}_{\phi 0} f(\tilde{x},\tilde{y},\tilde{z})   \hspace{0.2in}\mbox{with}\hspace{0.2in} \mathcal{E}_{\phi 0} = mc^2 \frac{I_p}{ I_A},
\label{efrz}
\eeq
for  $I_p=Nec/(\sqrt{2\pi} \sigma_z)$ being the peak current and   $I_A=e/(r_e c)$=17~kA the Alfven current, and
\beq
f(\tilde{x},\tilde{y}, \tilde{z}) = \int_0^\infty \frac{d\tau}{\sqrt{(\tau+\eta) (\tau+1)(\alpha \tau +1)}} 
\exp \left[- \frac{\tilde{x}^2}{2(\tau/\eta +1)}- \frac{\tilde{y}^2}{2(\tau+1)}- \frac{\tilde{z}^2}{2(\alpha \tau +1)} \right].
\label{frz1}
\eeq
For a  cylindrical beam, $\sigma_x=\sigma_y=\sigma_r$, and $\tilde{r}^2=\tilde{x}^2+\tilde{y}^2$.  Then $\mathcal{E}_{\phi}(x,y,z)$ becomes $\mathcal{E}_{\phi c}(r,z)$
\beq
\mathcal{E}_{\phi c}(r,z)=\mathcal{E}_{\phi 0} f_c(\tilde{r}, \tilde{z})
\vspace{-0.05in}
\eeq
for
\vspace{-0.05in}
\beq
f_c (r,z)=\int_{0}^{\infty} \frac{d\tau}{(1+\tau)\sqrt{1+\alpha \tau}}
\exp \left( -\frac{\tilde{r}^2}{2(1+\tau)} -\frac{\tilde{z}^2}{2(1+\alpha \tau)}\right).
\label{phic}
\vspace{0.1in}
\eeq
This expression is identical  to Eq.\,(2) of Ref.\,\cite{accl} with $\tau^{-1}=\lambda^2\sigma_{\perp}^2$.

\begin{figure}
\hspace{1.2in}
\includegraphics[scale=0.3]{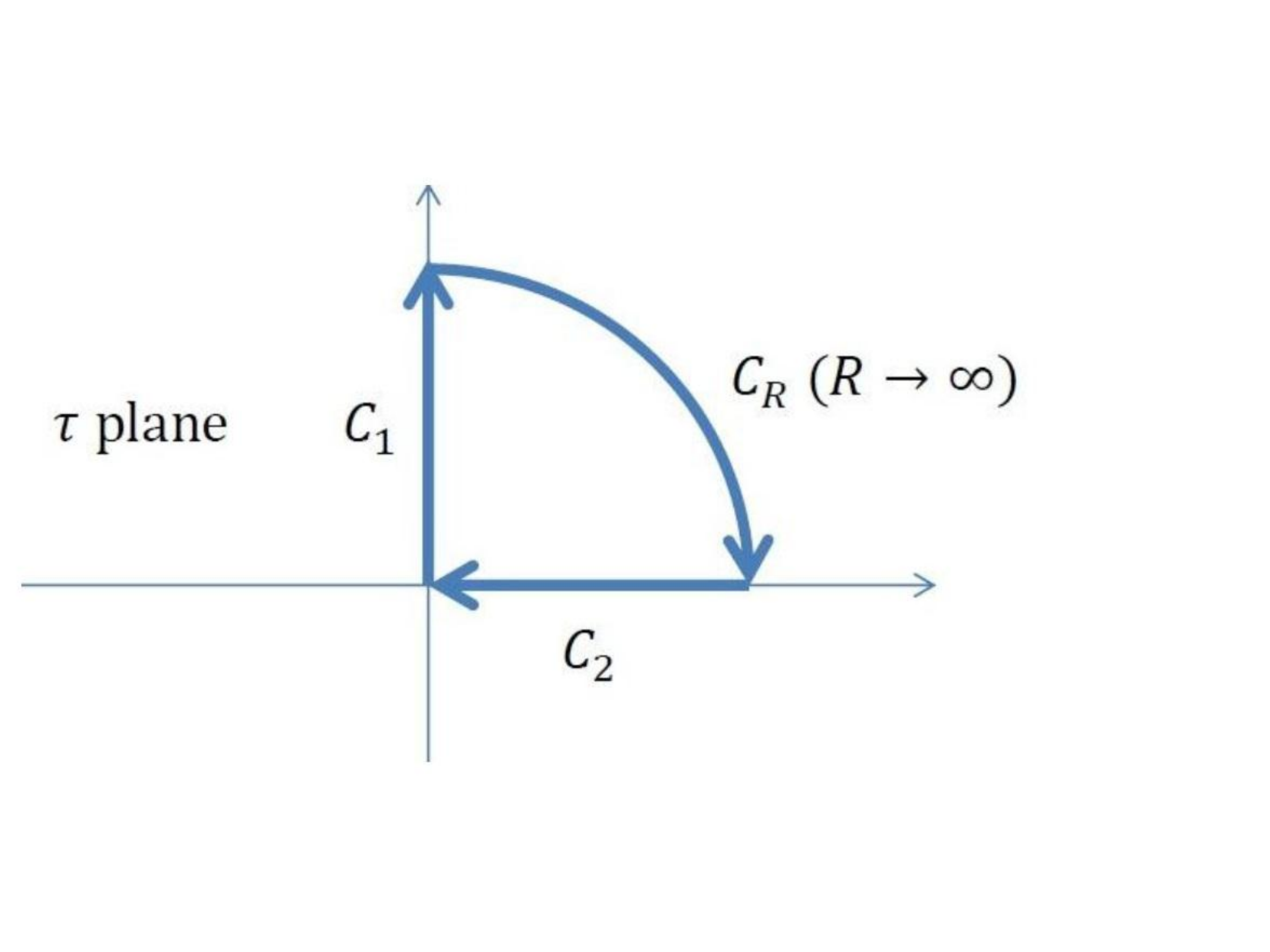}
\hspace{0.5in}
\vspace{-0.7in}
\caption{contour}
\label{contour} 
\end{figure}

For an infinitely  long bunch, $\alpha=0$ and thus the integral in Eq.~(\ref{phic}) diverges.  Even though this integral converges for  a  realistic beam of finite $\alpha$,  the  numerical integration converges very slowly as the upper limit approaches infinity. To increase the convergence rate for numerical computation, we change the variable from $\tau$ to $\kappa$ by $\kappa=\sinh^{-1}(\alpha \tau)$.   Eq.~(\ref{frz1}) becomes
\bea
f(\tilde{x},\tilde{y},\tilde{z}) &=& \int_0^{\infty} \frac{\cosh \kappa \, d\kappa}{\sqrt{(\alpha \eta +\sinh \kappa) (\alpha +\sinh \kappa) (1+\sinh \kappa)} }  \nonumber \\
&& \times \exp \left( -\frac{\alpha \,\tilde{x}^2}{2(\alpha +\sinh \kappa/\eta)}-\frac{\alpha \,\tilde{y}^2}{2(\alpha +\sinh \kappa)} -\frac{ \tilde{z}^2}{2(1 +\sinh \kappa)}  \right),
\label{fxyz}
\eea
where $f(\tilde{x},\tilde{y},\tilde{z})$ reaches its convergent result when the upper limit of integral $\kappa_{\mbox{\scriptsize max}} \sim 30$.  For the potential energy of particles in the  central slice of a  cylindrical bunch,  we set $\eta=1$ and $\tilde{z}=0$ in Eq.~(\ref{fxyz}). Using $U(\omega)$ to represent the  dependence of $f_c(r,0)$ in  Eq.~(\ref{phic}) on $\omega=\tilde{r}^2$, we get
\beq
U(\omega)= \int_0^{\infty} \frac{\cosh \kappa d\kappa}{(\alpha+\sinh \kappa)\sqrt{1+\sinh \kappa}} \exp \left(- \frac{\alpha\, \omega}{2(\alpha +\sinh \kappa)}
\right).
\label{UwNew}
\eeq
The behavior of $U(\omega)$ is shown in Fig.\,\ref{uw}. The form function for the probability of $\mathcal{E}_\phi$ over $\omega$ is
given by Eq.~(\ref{pefw}), where $dU/dw$ can be deduced from Eq.~(\ref{UwNew}).

\newpage 
\bibliography{apssamp}

\begin{thebibliography}{}
\bibitem{roll} Exception is found for the short duration of beam transport around roll-over compression \cite{lir08}.
\end{thebibliography}

\end{document}